\documentclass{iopart}
\usepackage{iopams}
\usepackage{graphicx}
\usepackage{dcolumn}
\usepackage{amsfonts}
\usepackage{latexsym}
\usepackage{amssymb}
\usepackage{verbatim}
\usepackage{amsthm}
\usepackage{wrapft}
\usepackage{mathrsfs}
\usepackage{mathptmx}

\newcommand{\RF}{{{\mathbb R}}}

\newtheorem{proposal}{Proposal}[section]

\newtheorem{conjecture}{Conjecture}[section]
\eqnobysec
\begin{document}
%%%%%%%%%%%%%%%%%%%%%%%%%%%%%%%%%%%%%%%%%%%%%%%%%%%%%%
%%%%%%%%%%%%%%%%%%%%%%%%%%%%%%%%%%%%%%%%%%%%%%%%%%%%%%
%%%%%%%%%%%%%%%%%%%%%%%%%%%%%%%%%%%%%%%%%%%%%%%%%%%%%%
%%%%%%%%%%%%%%%%%%%%%%%%%%%%%%%%%%%%%%%%%%%%%%%%%%%%%%
\title[
%Proposal of a gauge-invariant treatment of the $l=0,1$ mode
%perturbations ...
]{
  Proposal of a gauge-invariant treatment of $l=0,1$-mode
  perturbations on Schwarzschild background spacetime
}
%%%%%%%%%%%%%%%%%%%%%%%%%%%%%%%%%%%%%%%%%%%%%%%%%%%%%%
%%%%%%%%%%%%%%%%%%%%%%%%%%%%%%%%%%%%%%%%%%%%%%%%%%%%%%
%%%%%%%%%%%%%%%%%%%%%%%%%%%%%%%%%%%%%%%%%%%%%%%%%%%%%%
%%%%%%%%%%%%%%%%%%%%%%%%%%%%%%%%%%%%%%%%%%%%%%%%%%%%%%
\author{
  Kouji Nakamura
}
%%%%%%%%%%%%%%%%%%%%%%%%%%%%%%%%%%%%%%%%%%%%%%%%%%%%%%
%%%%%%%%%%%%%%%%%%%%%%%%%%%%%%%%%%%%%%%%%%%%%%%%%%%%%%
%%%%%%%%%%%%%%%%%%%%%%%%%%%%%%%%%%%%%%%%%%%%%%%%%%%%%%
%%%%%%%%%%%%%%%%%%%%%%%%%%%%%%%%%%%%%%%%%%%%%%%%%%%%%%
\address{
  Gravitational-Wave Science Project,
  National Astronomical Observatory of Japan,\\
  2-21-1, Osawa, Mitaka, Tokyo 181-8588, Japan
}
\ead{kouji.nakamura@nao.ac.jp}
%%%%%%%%%%%%%%%%%%%%%%%%%%%%%%%%%%%%%%%%%%%%%%%%%%%%%%
%%%%%%%%%%%%%%%%%%%%%%%%%%%%%%%%%%%%%%%%%%%%%%%%%%%%%%
%%%%%%%%%%%%%%%%%%%%%%%%%%%%%%%%%%%%%%%%%%%%%%%%%%%%%%
%%%%%%%%%%%%%%%%%%%%%%%%%%%%%%%%%%%%%%%%%%%%%%%%%%%%%%
\begin{abstract}
  A gauge-invariant treatment of the monopole- ($l=0$) and dipole
  ($l=1$) modes in linear perturbations of the Schwarzschild
  background spacetime is proposed.
  Through this gauge-invariant treatment, we derived the solutions to
  the linearized Einstein equation for these modes with a generic
  matter field.
  In the vacuum case, these solutions include the Kerr parameter
  perturbations in the $l=1$ odd modes and the additional mass
  parameter perturbations of the Schwarzschild mass in the $l=0$ even
  modes.
  The linearized version of Birkhoff's theorem is also confirmed in a
  gauge-invariant manner.
  In this sense, our proposal is reasonable.
\end{abstract}

%Uncomment for PACS numbers title message
%\pacs{04.20.-q, 04.20.Cv, 04.50.+h, 98.80.Jk}
% Keywords required only for MST, PB, PMB, PM, JOA, JOB?
%\vspace{2pc}
\noindent{\it Keywords}: gauge-invariant perturbations, Schwarzschild
spacetime, $l=0,1$ modes
% Uncomment for Submitted to journal title message
%\submitto{\CQG}
% Comment out if separate title page not required
%\maketitle

%%%%%%%%%%%%%%%%%%%%%%%%%%%%%%%%%%%%%%%%%%%%%%%%%%%%%%
%%%%%%%%%%%%%%%%%%%%%%%%%%%%%%%%%%%%%%%%%%%%%%%%%%%%%%
%%%%%%%%%%%%%%%%%%%%%%%%%%%%%%%%%%%%%%%%%%%%%%%%%%%%%%
%%%%%%%%%%%%%%%%%%%%%%%%%%%%%%%%%%%%%%%%%%%%%%%%%%%%%%
\section{Introduction}
\label{sec:introduction}
%%%%%%%%%%%%%%%%%%%%%%%%%%%%%%%%%%%%%%%%%%%%%%%%%%%%%%
%%%%%%%%%%%%%%%%%%%%%%%%%%%%%%%%%%%%%%%%%%%%%%%%%%%%%%
%%%%%%%%%%%%%%%%%%%%%%%%%%%%%%%%%%%%%%%%%%%%%%%%%%%%%%
%%%%%%%%%%%%%%%%%%%%%%%%%%%%%%%%%%%%%%%%%%%%%%%%%%%%%%

%*************************************************************

In 2015, the direct observation of gravitational waves was finally
accomplished by the Laser Interferometer Gravitational-wave
Observatory~\cite{LIGO-GW150914-2016}.
This event was the beginning of the gravitational-wave astronomy and
multi-messenger astronomy including
gravitational waves~\cite{LSC-homepage}.
One future direction of gravitational-wave astronomy
is the development as a precise science by the detailed studies of
source science and the tests of general-relativity through the
developments of the global network of gravitational-wave
detectors~\cite{LSC-homepage,Virgo-homepage}.
In addition to these ground-based detectors, some projects of space
gravitational-wave antenna are also
progressing~\cite{LISA-homepage,DECIGO-homepage}.
Among them, the Extreme-Mass-Ratio-Inspiral (EMRI) is one of
the targets of the Laser Interferometer Space
Antenna~\cite{LISA-homepage}.
The EMRI is a source of gravitational waves, which is the motion
of a stellar mass object around a supermassive black hole, and black
hole perturbation theories are used to describe the EMRI.
Therefore, theoretical sophistications of black hole perturbation
theories and their higher-order extensions are required to support the
development of precise experimental science.
This paper is motivated by such theoretical sophistication of black
hole perturbation theories toward higher-order perturbations.

%*************************************************************

Although realistic black holes have angular momentum and we must
consider the perturbation theory of a Kerr black hole for direct
application to the EMRI, further sophistication is possible even in
perturbation theories on the Schwarzschild background spacetime.
Based on the pioneering works by Regge and
Wheeler~\cite{T.Regge-J.A.Wheeler-1957} and
Zerilli~\cite{F.Zerilli-1970-PRL}, there have been many
studies on the perturbations in the Schwarzschild background
spacetime~\cite{V.Moncrief-1974a,U.H.Gerlach-U.K.Sengupta-1979a,C.Gundlach-J.M.Martin-Garcia-2000,D.Brizuela-J.M.Martin-Garcia-G.A.Mena-Marugan-2007}.
Because the Schwarzschild spacetime has a spherical symmetry, we
may decompose the perturbations on this spacetime using the spherical
harmonics $Y_{lm}$ and classify them into odd- and even-modes based on
their parity.
However, the current consensus is that $l=0$ and $l=1$ modes should be
separately treated~\cite{C.Gundlach-J.M.Martin-Garcia-2000}, and
``{\it gauge-invariant}'' treatments for $l=0$ and $l=1$ even-modes
remain unknown.

%*************************************************************

However, toward unambiguous sophisticated
nonlinear general-relativistic perturbation theories, we have been
developing the general formulation of a higher-order gauge-invariant
perturbation theory on a generic background
spacetime~\cite{K.Nakamura-2003,K.Nakamura-2005,K.Nakamura-2011,K.Nakamura-2014}
and have applied it to cosmological
perturbations~\cite{K.Nakamura-2006,K.Nakamura-2009a,K.Nakamura-2010}.
We briefly review our framework of the gauge-invariant perturbation
theory~\cite{K.Nakamura-2003,K.Nakamura-2005} in
Sec.~\ref{sec:review-of-perturbation-theroy}.
This framework is based on a conjecture (Conjecture
\ref{conjecture:decomposition-conjecture} below), which roughly
states that {\it we already know the procedure to find gauge-invariant
  variables for linear-order metric perturbations}.
A proof of Conjecture~\ref{conjecture:decomposition-conjecture} was
discussed in~\cite{K.Nakamura-2011}.
In this proof, we assumed the existence of the Green functions for
some elliptic derivative operators and ignored the kernel modes of these
elliptic derivative operators.
We call these kernel modes ``{\it zero modes},'' and the treatment
of these zero modes remains unclear.
We also called the problem of finding a treatment to these zero modes
as the``{\it zero-mode problem}.''

%*************************************************************

In the case of the gauge-invariant perturbation theory on the
Schwarzschild background spacetime, the zero modes are just above
the $l=0,1$ modes.
The special treatments of these modes become an obstacle when we
develop nonlinear perturbation theory because mode couplings owing to
the nonlinear effects produce higher-order $l=0,1$
modes~\cite{D.Brizuela-J.M.Martin-Garcia-G.A.Mena-Marugan-2007}.
Therefore, the finding of a gauge-invariant treatment of $l=0,1$ modes in the
perturbations on Schwarzschild background spacetime is a resolution of
the above zero-mode problem in a specific background spacetime.
Furthermore, this resolution is an important step of the development
of the higher-order gauge-invariant perturbation theory on the
Schwarzschild background spacetime.
In addition to the perturbation theory on a specific background
spacetime, this resolution will become a clue to the perturbation
theory on a generic background spacetime.

%*************************************************************

In this paper, we propose a gauge-invariant treatment of $l = 0,1$,
perturbations on the Schwarzschild background spacetime and show that
Conjecture~\ref{conjecture:decomposition-conjecture} is true even for
these modes.
We also derive the solutions to the linearized Einstein equation for
these modes.

%*************************************************************

The organization of this paper is as follows:
In Sec.~\ref{sec:review-of-perturbation-theroy}, we briefly review the
framework of the general-relativistic gauge-invariant perturbation
theory within the linear perturbation theory, although this framework
can be extended to higher-order perturbations~\cite{K.Nakamura-2003,K.Nakamura-2005,K.Nakamura-2011,K.Nakamura-2014};
In Sec.~\ref{sec:spherical_background_case}, we propose a strategy for
gauge-invariant treatments of $l=0,1$ modes after the explanation
of the situation, which in many studies requires the special
treatments of $l=0,1$ modes.
In Sec.~\ref{sec:Const._of_gauge_inv._var._linear_pert.}, we construct
gauge-invariant variables including $l=0,1$ modes through the proposal
described in Sec.~\ref{sec:spherical_background_case}.
This is a complete proof of
Conjecture~\ref{conjecture:decomposition-conjecture} for perturbations
on the background spacetimes with spherical symmetry.
In Sec.~\ref{sec:l=01_sol_Einstein_equations}, we derive the solutions
to the Einstein equations for $l=0,1$ modes.
Finally, in Sec.~\ref{sec:summary_and_discussion}, we provide some
concluding remarks and discussions regarding this research.

%*************************************************************

Throughout this paper, we use the unit $G=c=1$, where $G$ is Newton's
constant of gravitation, and $c$ is the velocity of light.

%*************************************************************

%%%%%%%%%%%%%%%%%%%%%%%%%%%%%%%%%%%%%%%%%%%%%%%%%%%%%%
%%%%%%%%%%%%%%%%%%%%%%%%%%%%%%%%%%%%%%%%%%%%%%%%%%%%%%
%%%%%%%%%%%%%%%%%%%%%%%%%%%%%%%%%%%%%%%%%%%%%%%%%%%%%%
\section{Brief review of general-relativistic gauge-invariant perturbation theory}
\label{sec:review-of-perturbation-theroy}
%%%%%%%%%%%%%%%%%%%%%%%%%%%%%%%%%%%%%%%%%%%%%%%%%%%%%%
%%%%%%%%%%%%%%%%%%%%%%%%%%%%%%%%%%%%%%%%%%%%%%%%%%%%%%
%%%%%%%%%%%%%%%%%%%%%%%%%%%%%%%%%%%%%%%%%%%%%%%%%%%%%%

%*********************************************************************

Herein, we briefly review the framework of the gauge-invariant
perturbation theory~\cite{K.Nakamura-2003,K.Nakamura-2005}.
In this review, we concentrate only on the linear perturbations,
because we treat only the linear perturbations within this paper.

%****************************************************************

In any perturbation theory, we always treat two spacetime manifolds.
One is the physical spacetime $({\cal M}_{{\rm ph}},\bar{g}_{ab})$,
which is identified with our nature itself, and we want to describe
this spacetime $({\cal M}_{{\rm ph}},\bar{g}_{ab})$ by perturbations.
The other is the background spacetime $({\cal M},g_{ab})$,
which is prepared as a reference by hand.
Note that these two spacetimes are distinct.
Furthermore, in any perturbation theory, we always write equations
for the perturbation of the variable $Q$ as follows:
\begin{equation}
  \label{eq:variable-symbolic-perturbation}
  Q(``p\mbox{''}) = Q_{0}(p) + \delta Q(p).
\end{equation}
Equation (\ref{eq:variable-symbolic-perturbation}) gives a
relation between variables on different manifolds.
Actually, $Q(``p\mbox{''})$ in
Eq.~(\ref{eq:variable-symbolic-perturbation}) is a variable on
${\cal M}_{\rm ph}$, whereas $Q_{0}(p)$ and $\delta Q(p)$ are variables
on ${\cal M}$.
Because we regard Eq.~(\ref{eq:variable-symbolic-perturbation}) as
a field equation, Eq.~(\ref{eq:variable-symbolic-perturbation})
includes an implicit assumption of the existence of a point
identification map ${\cal M}\rightarrow{\cal M}_{\rm ph}$ $:$
$p\in{\cal M}\mapsto ``p\mbox{''}\in{\cal M}_{\rm ph}$.
This identification map is a {\it gauge choice} in
general-relativistic perturbation theories.
This is the notion of the {\it second-kind gauge} pointed out by
Sachs~\cite{R.K.Sachs-1964}.
Note that this second-kind gauge is a different notion from the
degree of freedom of the coordinate transformation on a single
manifold, which is called the {\it first-kind
  gauge}~\cite{K.Nakamura-2010}.

%****************************************************************

To develop this understanding of the ``gauge,'' we introduce an
infinitesimal parameter $\lambda$ for perturbations and
$4+1$-dimensional manifold
${\cal N}={\cal M}_{{\rm ph}}\times\RF$ ($4=\dim{\cal M}$) such that
${\cal M}=\left.{\cal N}\right|_{\lambda=0}$ and
${\cal M}_{{\rm ph}}={\cal M}_{\lambda}=\left.{\cal N}\right|_{\RF=\lambda}$.
On ${\cal N}$, the gauge choice is regarded as a diffeomorphism
${\cal X}_{\lambda}:{\cal N}\rightarrow{\cal N}$ such that
${\cal X}_{\lambda}:{\cal M}\rightarrow{\cal M}_{\lambda}$.
This gauge choice is a point-identification.
Furthermore, we introduce a gauge choice ${\cal X}_{\lambda}$ as an
exponential map with a generator ${}^{{\cal X}}\!\eta^{a}$,
which is chosen such that its integral curve in ${\cal N}$ is
transverse to each ${\cal M}_{\lambda}$ everywhere on
${\cal N}$.
Points lying on the same integral curve are regarded as the
``same'' by the gauge choice ${\cal X}_{\lambda}$.

%****************************************************************

The first-order perturbation of the variable $Q$ on
${\cal M}_{\lambda}$ is defined as the pulled-back
${\cal X}_{\lambda}^{*}Q$ on ${\cal M}$, which is induced by
${\cal X}_{\lambda}$, and is expanded as
\begin{eqnarray}
  {\cal X}_{\lambda}^{*}Q
  =
  Q_{0}
  + \lambda \left.{\pounds}_{{}^{{\cal X}}\!\eta}Q\right|_{{\cal M}}
  + O(\lambda^{2}),
  \label{eq:perturbative-expansion-of-Q-def}
\end{eqnarray}
where $Q_{0}=\left.Q\right|_{{\cal M}}$ is the background value of
$Q$ and all terms in
Eq.~(\ref{eq:perturbative-expansion-of-Q-def}) are evaluated on the
background spacetime ${\cal M}$.
Because Eq.~(\ref{eq:perturbative-expansion-of-Q-def}) is the
perturbative expansion of ${\cal X}^{*}_{\lambda}Q_{\lambda}$,
the first-order perturbation of $Q$ is given by
${}^{(1)}_{{\cal X}}\!Q:=\left.{\pounds}_{{}^{{\cal X}}\!\eta}Q\right|_{{\cal M}}$.

%****************************************************************

When we have two gauge choices ${\cal X}_{\lambda}$ and
${\cal Y}_{\lambda}$ with the generators ${}^{{\cal X}}\!\eta^{a}$
and ${}^{{\cal Y}}\!\eta^{a}$, respectively, and when these
generators have different tangential components
to each ${\cal M}_{\lambda}$, ${\cal X}_{\lambda}$ and
${\cal Y}_{\lambda}$ are regarded as {\it different gauge choices}.
A {\it gauge-transformation} is regarded as the change of the
point-identification
${\cal X}_{\lambda}\rightarrow{\cal Y}_{\lambda}$, which is given by
the diffeomorphism
$\Phi_{\lambda}$ $:=$
$\left({\cal X}_{\lambda}\right)^{-1}\circ{\cal Y}_{\lambda}$
$:$ ${\cal M}$ $\rightarrow$ ${\cal M}$.
The diffeomorphism $\Phi_{\lambda}$ does change the
point-identification.
Here, $\Phi_{\lambda}$ induces a pull-back from the representation
${\cal X}_{\lambda}^{*}\!Q_{\lambda}$ to the representation
${\cal Y}_{\lambda}^{*}\!Q_{\lambda}$ as
${\cal Y}_{\lambda}^{*}\!Q_{\lambda}=\Phi_{\lambda}^{*}{\cal X}_{\lambda}^{*}\!Q_{\lambda}$.
From general arguments of the Taylor
expansion\cite{M.Bruni-S.Matarrese-S.Mollerach-S.Sonego-1997}, the
pull-back $\Phi_{\lambda}^{*}$ is expanded as
\begin{eqnarray}
  {\cal Y}_{\lambda}^{*}\!Q_{\lambda}
  &=&
  {\cal X}_{\lambda}^{*}\!Q_{\lambda}
  + \lambda {\pounds}_{\xi_{(1)}} {\cal X}_{\lambda}^{*}\!Q_{\lambda}
  + O(\lambda^{2}),
  \label{eq:Bruni-46-one}
\end{eqnarray}
where $\xi_{(1)}^{a}$ is the generator of $\Phi_{\lambda}$.
From Eqs.~(\ref{eq:perturbative-expansion-of-Q-def}) and
(\ref{eq:Bruni-46-one}), each order gauge-transformation is
given as
\begin{eqnarray}
  \label{eq:Bruni-47-one}
  {}^{(1)}_{\;{\cal Y}}\!Q - {}^{(1)}_{\;{\cal X}}\!Q &=&
  {\pounds}_{\xi_{(1)}}Q_{0}.
\end{eqnarray}
We also employ the {\it order by order gauge invariance} as a
concept of gauge invariance\cite{K.Nakamura-2009a}.
We call the $k$th-order perturbation ${}^{(k)}_{{\cal X}}\!Q$ as
gauge invariant if and only if
${}^{(k)}_{\;\cal X}\!Q = {}^{(k)}_{\;\cal Y}\!Q$ for any gauge choice
${\cal X}_{\lambda}$ and ${\cal Y}_{\lambda}$.

%****************************************************************

Based on the above setup, we proposed a procedure to construct
gauge-invariant variables of higher-order
perturbations\cite{K.Nakamura-2003,K.Nakamura-2005}.
In this paper, we concentrate only on the explanations of the linear
perturbations.
First, we expand the metric on the physical spacetime
${\cal M}_{\lambda}$, which was pulled back to the background
spacetime ${\cal M}$ through a gauge choice ${\cal X}_{\lambda}$ as
\begin{eqnarray}
  {\cal X}^{*}_{\lambda}\bar{g}_{ab}
  &=&
  g_{ab} + \lambda {}_{{\cal X}}\!h_{ab}
  + O(\lambda^{2}).
  \label{eq:metric-expansion}
\end{eqnarray}
Although the expression (\ref{eq:metric-expansion}) depends
entirely on the gauge choice ${\cal X}_{\lambda}$, henceforth,
we do not explicitly express the index of the gauge choice
${\cal X}_{\lambda}$ in the expression if there is no
possibility of confusion.
The important premise of our proposal was the following
conjecture\cite{K.Nakamura-2003,K.Nakamura-2005} for the linear metric
perturbation $h_{ab}$:
\begin{conjecture}
  \label{conjecture:decomposition-conjecture}
  If the gauge-transformation rule for a tensor field $h_{ab}$ is
  given by ${}_{{\cal Y}}\!h_{ab}$ $-$ ${}_{{\cal X}}\!h_{ab}$ $=$
  ${\pounds}_{\xi_{(1)}}g_{ab}$ with the background metric $g_{ab}$,
  there then exist a tensor field ${\cal F}_{ab}$ and a vector
  field $Y^{a}$ such that $h_{ab}$ is decomposed as $h_{ab}$ $=:$
  ${\cal F}_{ab}$ $+$ ${\pounds}_{Y}g_{ab}$, where ${\cal F}_{ab}$ and
  $Y^{a}$ are transformed into ${}_{{\cal Y}}\!{\cal F}_{ab}$ $-$
  ${}_{{\cal X}}\!{\cal F}_{ab}$ $=$ $0$ and ${}_{{\cal Y}}\!Y^{a}$
  $-$ ${}_{{\cal X}}\!Y^{a}$ $=$ $\xi^{a}_{(1)}$ under the gauge
  transformation, respectively.
\end{conjecture}
We call ${\cal F}_{ab}$ and $Y^{a}$ as the
{\it gauge-invariant} and {\it gauge-variant} parts
of $h_{ab}$, respectively.

%*************************************************************

The proof of Conjecture~\ref{conjecture:decomposition-conjecture} is
highly nontrivial~\cite{K.Nakamura-2011}, and it was
found that the gauge-invariant variables are essentially non-local, as
mentioned in Sec.~\ref{sec:introduction}.
Despite this non-triviality, once we accept
Conjecture~\ref{conjecture:decomposition-conjecture},
we can construct gauge-invariant variables for the linear perturbation
of an arbitrary tensor field ${}_{\cal X}^{(1)}\!Q$, whose
gauge-transformation is given by Eq.~(\ref{eq:Bruni-47-one}), through
the gauge-variant part of the metric perturbation $Y_{a}$ in
Conjecture~\ref{conjecture:decomposition-conjecture} as
\begin{eqnarray}
  \label{eq:gauge-inv-Q-def}
  {}^{(1)}\!{\cal Q} := {}_{\cal X}^{(1)}\!Q - {\pounds}_{{}_{\cal X}\!Y}Q_{0}.
\end{eqnarray}
This definition implies that the linear perturbation
${}_{\cal X}^{(1)}\!Q$ of an arbitrary tensor field
${\cal X}_{\lambda}^{*}Q$ is always decomposed into its gauge-invariant
part ${}^{(1)}\!{\cal Q}$ and gauge-variant part
${\pounds}_{{}_{\cal X}\!Y}Q_{0}$ as
\begin{eqnarray}
  \label{eq:arbitrary-Q-decomp}
  {}_{\cal X}^{(1)}\!Q = {}^{(1)}\!{\cal Q} + {\pounds}_{{}_{\cal X}\!Y}Q_{0}.
\end{eqnarray}
As an example, the linearized Einstein tensor
${}_{\cal X}^{(1)}G_{a}^{\;\;b}$ and the linear perturbation of the
energy-momentum tensor ${}_{\cal X}^{(1)}T_{a}^{\;\;b}$ are also
decomposed as
\begin{eqnarray}
  \label{eq:Gab-Tab-decomp}
  {}_{\cal X}^{(1)}\!G_{a}^{\;\;b}
  =
  {}^{(1)}\!{\cal G}_{a}^{\;\;b}\left[{\cal F}\right] + {\pounds}_{{}_{\cal X}\!Y}G_{a}^{\;\;b}
  ,
  \quad
  {}_{\cal X}^{(1)}\!T_{a}^{\;\;b}
  =
  {}^{(1)}\!{\cal T}_{a}^{\;\;b} + {\pounds}_{{}_{\cal X}\!Y}T_{a}^{\;\;b}
  ,
\end{eqnarray}
where $G_{ab}$ and $T_{ab}$ are the background values of the Einstein
tensor and the energy-momentum tensor, respectively.
The gauge-invariant part ${}^{(1)}\!{\cal G}_{a}^{\;\;b}$ of the
linear-order perturbation of the Einstein tensor is given by
\begin{eqnarray}
  \label{eq:linear-Einstein-AIA2010-2}
  \!\!\!\!\!\!\!\!\!\!\!\!\!\!\!\!
  &&
     {}^{(1)}{\cal G}_{a}^{\;\;b}\left[A\right]
     :=
     {}^{(1)}\Sigma_{a}^{\;\;b}\left[A\right]
     - \frac{1}{2} \delta_{a}^{\;\;b} {}^{(1)}\Sigma_{c}^{\;\;c}\left[A\right]
     ,
  \\
  \label{eq:(1)Sigma-def-linear}
  \!\!\!\!\!\!\!\!\!\!\!\!\!\!\!\!
  &&
     {}^{(1)}\Sigma_{a}^{\;\;b}\left[A\right]
     :=
     - 2 \nabla_{[a}^{}H_{d]}^{\;\;\;bd}\left[A\right]
     - A^{cb} R_{ac}
     , \quad
     H_{ba}^{\;\;\;\;c}\left[A\right]
     :=
     \nabla_{(a}A_{b)}^{\;\;\;\;c} - \frac{1}{2} \nabla^{c}A_{ab}
     ,
\end{eqnarray}
where $A_{ab}$ is an arbitrary tensor field of the second
rank and $\Sigma_{a}^{\;\;b}\left[A\right]$ corresponds to the
gauge-invariant part of the linear perturbation of the Ricci tensor $R_{a}^{\;\;b}$~\cite{K.Nakamura-2005,K.Nakamura-2010}.
Then, using the background Einstein equation
$G_{a}^{\;\;b}=8\pi T_{a}^{\;\;b}$, the linearized Einstein equation
${}_{\cal X}^{(1)}\!G_{ab}=8\pi{}_{\cal X}^{(1)}\!T_{ab}$ is
automatically given in the gauge-invariant form
\begin{eqnarray}
  \label{eq:einstein-equation-gauge-inv}
  {}^{(1)}\!{\cal G}_{ab}\left[{\cal F}\right] = 8 \pi {}^{(1)}\!{\cal T}_{ab}
\end{eqnarray}
even if the background Einstein equation $G_{ab}=8\pi T_{ab}$ is
nontrivial.
We also note that, in the case of a vacuum background case, i.e.,
$8\pi T_{ab}=G_{ab} = 0$, Eq.~(\ref{eq:Gab-Tab-decomp}) shows that the
linear perturbations of the Einstein tensor and the energy-momentum
tensor is automatically gauge-invariant.

%*************************************************************

We can also derive the perturbation of the divergence of
$\bar{\nabla}_{a}\bar{T}_{b}^{\;\;a}$ of the second-rank tensor
$\bar{T}_{b}^{\;\;a}$ on $({\cal M}_{\rm ph},\bar{g}_{ab})$.
Through the gauge choice ${\cal X}_{\lambda}$, $\bar{T}_{b}^{\;\;a}$
is pulled back to ${\cal X}_{\lambda}^{*}\bar{T}_{b}^{\;\;a}$ on the
background spacetime $({\cal M},g_{ab})$, and the
covariant derivative operator $\bar{\nabla}_{a}$ on
$({\cal M}_{\rm ph},\bar{g}_{ab})$ is pulled back to a derivative
operator
$\bar{\nabla}_{a}(={\cal X}_{\lambda}^{*}\bar{\nabla}_{a}({\cal X}_{\lambda}^{-1})^{*})$
on $({\cal M},g_{ab})$.
Note that the derivative $\bar{\nabla}_{a}$ is the covariant
derivative associated with the metric
${\cal X}_{\lambda}\bar{g}_{ab}$, whereas the derivative $\nabla_{a}$ on
the background spacetime $({\cal M},g_{ab})$ is the covariant derivative
associated with the background metric $g_{ab}$.
Bearing in mind the difference in these derivative, the first-order
perturbation of $\bar{\nabla}_{a}\bar{T}_{b}^{\;\;a}$ is given by
\begin{eqnarray}
  \label{eq:linear-perturbation-of-div-Tab}
  {}^{(1)}\!\left(\bar{\nabla}_{a}\bar{T}_{b}^{\;\;a}\right)
  =
  \nabla_{a}{}^{(1)}\!{\cal T}_{b}^{\;\;a}
  +
  H_{ca}^{\;\;\;\;a}\left[{\cal F}\right] T_{b}^{\;\;c}
  -
  H_{ba}^{\;\;\;\;c}\left[{\cal F}\right] T_{c}^{\;\;a}
  +
  {\pounds}_{Y}\nabla_{a}T_{b}^{\;\;a}
  .
\end{eqnarray}
The derivation of the formula
(\ref{eq:linear-perturbation-of-div-Tab}) is given in
Ref.~\cite{K.Nakamura-2005}.
If the tensor field $\bar{T}_{b}^{\;\;a}$ is the Einstein tensor
$\bar{G}_{a}^{\;\;b}$, Eq.~(\ref{eq:linear-perturbation-of-div-Tab})
yields the linear-order perturbation of the Bianchi identity
\begin{eqnarray}
  \label{eq:linear-perturbation-of-div-Gab}
  \nabla_{a}{}^{(1)}\!{\cal G}_{b}^{\;\;a}\left[{\cal F}\right]
  +
  H_{ca}^{\;\;\;\;a}\left[{\cal F}\right] G_{b}^{\;\;c}
  -
  H_{ba}^{\;\;\;\;c}\left[{\cal F}\right] G_{c}^{\;\;a}
  =
  0
\end{eqnarray}
and if the background Einstein tensor vanishes $G_{a}^{\;\;b}=0$, we
obtain the identity
\begin{eqnarray}
  \label{eq:linear-perturbation-of-div-Gab-vacuum}
  \nabla_{a}{}^{(1)}\!{\cal G}_{b}^{\;\;a}\left[{\cal F}\right]
  =
  0.
\end{eqnarray}
By contrast, if the tensor field $\bar{T}_{b}^{\;\;a}$ is the
energy-momentum tensor, Eq.~(\ref{eq:linear-perturbation-of-div-Tab})
yields the continuity equation of the energy-momentum tensor
\begin{eqnarray}
  \label{eq:linear-perturbation-of-div-Tab-ene-mon}
  \nabla_{a}{}^{(1)}\!{\cal T}_{b}^{\;\;a}
  +
  H_{ca}^{\;\;\;\;a}\left[{\cal F}\right] T_{b}^{\;\;c}
  -
  H_{ba}^{\;\;\;\;c}\left[{\cal F}\right] T_{c}^{\;\;a}
  =
  0
  ,
\end{eqnarray}
where we used the background continuity equation
$\nabla_{a}T_{b}^{\;\;a}=0$.
If the background spacetime is vacuum $T_{ab}=0$,
Eq.~(\ref{eq:linear-perturbation-of-div-Tab-ene-mon}) yields
a linear perturbation of the energy-momentum tensor given by
\begin{eqnarray}
  \label{eq:divergence-barTab-linear-vac-back-u}
  \nabla_{a}{}^{(1)}\!{\cal T}_{b}^{\;\;a}
  =
  0
  .
\end{eqnarray}

%*************************************************************

Thus, starting from the
Conjecture~\ref{conjecture:decomposition-conjecture}, we can develop
the gauge-invariant perturbation theory through the above framework.
Furthermore, this formulation can be extended to any order
perturbations~\cite{K.Nakamura-2003,K.Nakamura-2005,K.Nakamura-2011,K.Nakamura-2014}
from Conjecture~\ref{conjecture:decomposition-conjecture}.
In this sense, the proof of the
Conjecture~\ref{conjecture:decomposition-conjecture} is crucial to
this framework.
We should note that the decomposition of the metric perturbation
$h_{ab}$ into its gauge-invariant part ${\cal F}_{ab}$ and
into its gauge-variant part $Y^{a}$ is not
unique~\cite{K.Nakamura-2009a,K.Nakamura-2010}.
For example, The gauge-invariant part ${\cal F}_{ab}$ has six components and we
can create the gauge-invariant vector field $Z^{a}$ through the
component ${\cal F}_{ab}$ such that the gauge-transformation of the
vector field $Z^{a}$ is given by ${}_{{\cal Y}}\!Z^{a}$ $-$
${}_{{\cal X}}\!Z^{a}$ $=$ $0$.
Using this gauge-invariant vector field $Z^{a}$, the original metric
perturbation can be expressed as follows:
\begin{eqnarray}
  \label{eq:gauge-inv-nonunique}
  h_{ab}
  =
  {\cal F}_{ab} - {\pounds}_{Z}g_{ab}
  + {\pounds}_{Z+Y}g_{ab}
  =:
  {\cal H}_{ab} + {\pounds}_{X}g_{ab}
  .
\end{eqnarray}
The tensor field ${\cal H}_{ab}:={\cal F}_{ab} - {\pounds}_{Z}g_{ab}$
is also regarded as the gauge-invariant part of the perturbation
$h_{ab}$ because ${}_{\cal Y}\!{\cal H}_{ab}-{}_{\cal X}\!{\cal
  H}_{ab}=0$.
Similarly, the vector field $X^{a}:=Z^{a}+Y^{a}$ is also regarded as
the gauge-variant part of the perturbation $h_{ab}$ because
${}_{{\cal Y}}\!X^{a}$ $-$ ${}_{{\cal X}}\!X^{a}$ $=$ $\xi^{a}_{(1)}$.
This non-uniqueness appears in the solutions derived in
Sec.~\ref{sec:l=01_sol_Einstein_equations}.

%****************************************************************

Finally, we comment on the relation between the gauge-transformation
$\Phi_{\lambda}$ and the coordinate transformation~\cite{K.Nakamura-2010}.
As mentioned above, the notion of the second-kind gauges above is
different from the notion of the degree of freedom of the coordinate
transformation on a single manifold.
However, the gauge-transformation $\Phi_{\lambda}$ of the second kind
induces the coordinate transformations.
To see this, we introduce the coordinate system
$\{O_{\alpha},\psi_{\alpha}\}$ on the background spacetime ${\cal M}$,
where $O_{\alpha}$ are open sets on the background spacetime and
$\psi_{\alpha}$ are diffeomorphisms from $O_{\alpha}$ to $\RF^{4}$
($4=\dim{\cal M}$).
The coordinate system $\{O_{\alpha},\psi_{\alpha}\}$ is the set of 
collections of the pair of open sets $O_{\alpha}$ and diffeomorphism
$O_{\alpha}\mapsto\RF^{4}$.
If we employ a gauge choice ${\cal X}_{\lambda}$ of the second kind,
we have the correspondence of the physical spacetime
${\cal M}_{\lambda}={\cal M}_{{\rm ph}}$ and the background spacetime
${\cal M}$.
Together with the coordinate system $\psi_{\alpha}$ on ${\cal M}$,
this correspondence between ${\cal M}_{\lambda}$ and ${\cal M}$
induces the coordinate system on ${\cal M}_{\lambda}$.
Actually, ${\cal X}_{\lambda}(O_{\alpha})$ for each $\alpha$ is an
open set of ${\cal M}_{\lambda}$.
Then, $\psi_{\alpha}\circ{\cal X}_{\lambda}^{-1}$ becomes a
diffeomorphism from an open set ${\cal X}_{\lambda}(O_{\alpha})\subset
{\cal M}_{\lambda}$ to $\RF^{4}$.
This diffeomorphism $\psi_{\alpha}\circ{\cal X}_{\lambda}^{-1}$
induces a coordinate system of an open set on ${\cal M}_{\lambda}$.
When we have two different gauge choices ${\cal X}_{\lambda}$ and
${\cal Y}_{\lambda}$ of the second kind,
$\psi_{\alpha}\circ{\cal X}_{\lambda}^{-1}$ $:$
${\cal M}_{\lambda}$ $\mapsto$ $\RF^{4}$ ($\{x^{\mu}\}$) and
$\psi_{\alpha}\circ{\cal Y}_{\lambda}^{-1}$ $:$ ${\cal M}_{\lambda}$
$\mapsto$ $\RF^{4}$ ($\{y^{\mu}\}$) become different coordinate systems
on ${\cal M}_{\lambda}$.
We can also consider the coordinate transformation from the coordinate
system $\psi_{\alpha}\circ{\cal X}_{\lambda}^{-1}$ to another
coordinate system $\psi_{\alpha}\circ{\cal Y}_{\lambda}^{-1}$.
Because the gauge transformation
${\cal X}_{\lambda}\rightarrow{\cal Y}_{\lambda}$ is induced by the
diffeomorphism $\Phi_{\lambda}$ $:=$
$\left({\cal X}_{\lambda}\right)^{-1}\circ{\cal Y}_{\lambda}$, this
diffeomorphism $\Phi_{\lambda}$ induces the coordinate transformation
as
\begin{eqnarray}
  \label{eq:induced-coordinate-trans}
  y^{\mu}(q) := x^{\mu}(p) = \left((\Phi_{\lambda}^{-1})^{*}x^{\mu}\right)(q)
\end{eqnarray}
in the passive point of
view~\cite{K.Nakamura-2003,M.Bruni-S.Matarrese-S.Mollerach-S.Sonego-1997},
where $p$, $q$ $\in$ ${\cal M}$ are identified to the same point
$``p\mbox{"}$ $\in$ ${\cal M}_{\lambda}$ by the gauge choices
${\cal X}_{\lambda}$ and ${\cal Y}_{\lambda}$, respectively.
If we represent this coordinate transformation in terms of the Taylor
expansion (\ref{eq:Bruni-46-one}), we have the coordinate transformation
\begin{eqnarray}
  \label{eq:infinitesimal-coordinate-trans-explicit}
  y^{\mu}(q) = x^{\mu}(q) - \lambda \xi^{\mu}_{(1)}(q) + O(\lambda^{2}).
\end{eqnarray}
We should emphasize that the coordinate transformation
(\ref{eq:infinitesimal-coordinate-trans-explicit}) is not the starting
point of the gauge-transformation but a result of the above framework.
Because our above framework of the gauge-invariant perturbation theory
is constructed without a coordinate transformation
(\ref{eq:infinitesimal-coordinate-trans-explicit}), we avoid the use of the
coordinate transformation
(\ref{eq:infinitesimal-coordinate-trans-explicit}) as much as possible.

%****************************************************************

%%%%%%%%%%%%%%%%%%%%%%%%%%%%%%%%%%%%%%%%%%%%%%%%%%%%%%
%%%%%%%%%%%%%%%%%%%%%%%%%%%%%%%%%%%%%%%%%%%%%%%%%%%%%%
%%%%%%%%%%%%%%%%%%%%%%%%%%%%%%%%%%%%%%%%%%%%%%%%%%%%%%
\section{Linear perturbations on spherically symmetric background}
\label{sec:spherical_background_case}
%%%%%%%%%%%%%%%%%%%%%%%%%%%%%%%%%%%%%%%%%%%%%%%%%%%%%%
%%%%%%%%%%%%%%%%%%%%%%%%%%%%%%%%%%%%%%%%%%%%%%%%%%%%%%
%%%%%%%%%%%%%%%%%%%%%%%%%%%%%%%%%%%%%%%%%%%%%%%%%%%%%%

%*********************************************************************

Here, we use the 2+2 formulation of the perturbations on spherically
symmetric background spacetimes, which was originally proposed by
Gerlach and Sengupta~\cite{U.H.Gerlach-U.K.Sengupta-1979a}.
The topological space of spherically symmetric spacetimes is the
direct product ${\cal M}={\cal M}_{1}\times S^{2}$, and the metric on
this spacetime is
\begin{eqnarray}
  \label{eq:background-metric-2+2}
  g_{ab}
  &=&
      y_{ab} + r^{2}\gamma_{ab}
      ,
      \;\;\;
      y_{ab} = y_{AB} (dx^{A})_{a}(dx^{B})_{b}
      ,
      \;\;\;
      \gamma_{ab} = \gamma_{pq} (dx^{p})_{a} (dx^{q})_{b}
      ,
\end{eqnarray}
where $x^{A} = (t,r)$ and $x^{p}=(\theta,\phi)$.
In addition, $\gamma_{pq}$ is a metric of the unit sphere.
In the Schwarzschild spacetime, the metric
(\ref{eq:background-metric-2+2}) is given by
\begin{eqnarray}
  \label{eq:background-metric-2+2-y-comp-Schwarzschild}
  y_{ab}
  &=&
      - f (dt)_{a}(dt)_{b}
      +
      f^{-1} (dr)_{a}(dr)_{b}
      ,
      \quad
      f = 1 - \frac{2M}{r}
      ,
  \\
  \label{eq:background-metric-2+2-gamma-comp-Schwarzschild}
  \gamma_{ab}
  &=&
  (d\theta)_{a}(d\theta)_{b}
  +
  \sin^{2}\theta(d\phi)_{a}(d\phi)_{b}
  .
\end{eqnarray}
On this background spacetime $({\cal M},g_{ab})$, we consider the
components of the metric perturbation as
\begin{eqnarray}
  h_{ab}
  =
  h_{AB} (dx^{A})_{a}(dx^{B})_{b}
  +
  2 h_{Ap} (dx^{A})_{(a}(dx^{p})_{b)}
  +
  h_{pq} (dx^{p})_{a}(dx^{q})_{b}
  .
\end{eqnarray}
In~\cite{U.H.Gerlach-U.K.Sengupta-1979a},
these components of the metric perturbation are decomposed through the
decomposition~\cite{J.W.York-1973}
using the spherical harmonics $S = Y_{lm}$ as follows:
\begin{eqnarray}
  \label{eq:hAB-fourier}
  \!\!\!\!\!\!\!\!\!\!\!\!\!\!\!\!\!\!\!\!\!\!\!\!\!\!\!\!\!\!\!\!\!\!\!\!\!\!\!\!\!\!\!\!\!\!\!\!
  h_{AB}
  &=&
      \sum_{l,m} \tilde{h}_{AB} S
      ,
  \quad
      % \label{eq:hAp-fourier}
      h_{Ap}
      =
      r \sum_{l,m} \left[
      \tilde{h}_{(e1)A} \hat{D}_{p}S
      +
      \tilde{h}_{(o1)A} \epsilon_{pq} \hat{D}^{q}S
      \right]
      ,
  \\
  \label{eq:hpq-fourier}
  \!\!\!\!\!\!\!\!\!\!\!\!\!\!\!\!\!\!\!\!\!\!\!\!\!\!\!\!\!\!\!\!\!\!\!\!\!\!\!\!\!\!\!\!\!\!\!\!
  h_{pq}
  &=&
      r^{2} \sum_{l,m} \left[
      \frac{1}{2} \gamma_{pq} \tilde{h}_{(e0)} S
      +
      \tilde{h}_{(e2)} \left(
      \hat{D}_{p}\hat{D}_{q} - \frac{1}{2} \gamma_{pq} \hat{\Delta}
      \right) S
%      \right.
%      \nonumber\\
%  && \quad\quad\quad
%     \left.
      +
      2 \tilde{h}_{(o2)} \epsilon_{r(p} \hat{D}_{q)}\hat{D}^{r} S
      \right]
      ,
\end{eqnarray}
where $\hat{D}_{p}$ is the covariant derivative associated with
the metric $\gamma_{pq}$ on $S^{2}$,
$\hat{D}^{p}:=\gamma^{pq}\hat{D}_{q}$, and
$\epsilon_{pq}=\epsilon_{[pq]}$ is the totally antisymmetric
tensor on $S^{2}$.
Although the matrix representations of the independent harmonic
functions are used in pioneer
works~\cite{T.Regge-J.A.Wheeler-1957,F.Zerilli-1970-PRL},
these are equivalent to the covariant form
(\ref{eq:hAB-fourier})--(\ref{eq:hpq-fourier}).
This decomposition formulae
(\ref{eq:hAB-fourier})--(\ref{eq:hpq-fourier}) is the starting point
of the 2 + 2 formulation of the perturbations on spherically symmetric
spacetimes proposed by Gerlach and
Sengupta~\cite{U.H.Gerlach-U.K.Sengupta-1979a}.
Gerlach and Sengupta showed the constructions of gauge-invariant
variables for $l\geq 2$ and derived the linearized Einstein
equations.
However, separate treatments are required for $l = 0,1$
modes~\cite{C.Gundlach-J.M.Martin-Garcia-2000}, which is the main
target of this paper.

%*********************************************************************

Herein, we describe the situation of $l = 0,1$ modes in the 2 + 2
formulation.
Note that the decompositions
(\ref{eq:hAB-fourier})--(\ref{eq:hpq-fourier}) implicitly state that
the Green function of the
derivative operators $\hat{\Delta}:=\hat{D}^{r}\hat{D}_{r}$ and
$\hat{\Delta}+2:=\hat{D}^{r}\hat{D}_{r}+2$ should exist if the
one-to-one correspondence between $\{h_{Ap},$ $h_{pq}\}$ and
$\{\tilde{h}_{(e1)A},$ $\tilde{h}_{(o1)A},$ $\tilde{h}_{(e0)},$
$\tilde{h}_{(e2)},$ $\tilde{h}_{(o2)}\}$ is guaranteed.
Because the eigenvalue of the derivative operator $\hat{\Delta}$ on
$S^{2}$ is $-l(l+1)$, the kernels of the operators $\hat{\Delta}$ and
$\hat{\Delta}+2$ are $l = 0$ and $l = 1$ modes, respectively.
Thus, the one-to-one correspondence between
$\{h_{Ap},$ $h_{pq}\}$ and
$\{\tilde{h}_{(e1)A},$ $\tilde{h}_{(o1)A},$ $\tilde{h}_{(e0)},$
$\tilde{h}_{(e2)},$ $\tilde{h}_{(o2)}\}$ is lost for $l = 0,1$ modes in
decomposition formulae
(\ref{eq:hAB-fourier})--(\ref{eq:hpq-fourier}) with $S=Y_{lm}$.
If we choose the decomposition formulae
(\ref{eq:hAB-fourier})--(\ref{eq:hpq-fourier}) as the starting point
of the metric perturbations even for $l = 0,1$ modes, the
gauge-invariance of $l = 0,1$-mode perturbations becomes unclear owing
to the loss of this one-to-one correspondence.
Therefore, we should regard that the second decomposition
formula in Eq.~(\ref{eq:hAB-fourier}) with $S=Y_{lm}$ does not
include the $l = 0$ mode and the formula (\ref{eq:hpq-fourier}) with
$S=Y_{lm}$ does not include $l = 0,1$ modes.
This situation is also seen from the harmonics
$\hat{D}_{p}Y_{lm}$, $\epsilon_{pr}\hat{D}^{r}Y_{lm}$,
$\left(\hat{D}_{p}\hat{D}_{q}-\frac{1}{2}\gamma_{pq}\hat{\Delta}\right)Y_{lm}$,
and $2\epsilon_{r(p}\hat{D}_{q)}\hat{D}^{r}Y_{lm}$ for the $l=0,1$
modes as
$\hat{D}_{p}Y_{00}$ $=$ $\epsilon_{pr}\hat{D}^{r}Y_{00}$ $=$ $0$,
$\left(\hat{D}_{p}\hat{D}_{q}-\frac{1}{2}\gamma_{pq}\hat{\Delta}\right)Y_{00}$
$=$ $2\epsilon_{r(p}\hat{D}_{q)}\hat{D}^{r}Y_{00}$ $=$ $0$, and
$\left(\hat{D}_{p}\hat{D}_{q}-\frac{1}{2}\gamma_{pq}\hat{\Delta}\right)Y_{1m}$
$=$ $2\epsilon_{r(p}\hat{D}_{q)}\hat{D}^{r}Y_{1m}$ $=$ $0$.
Then, for these kernel modes, the one-to-one correspondence between
$\{h_{Ap},$ $h_{pq}\}$ and
$\{\tilde{h}_{(e1)A},$ $\tilde{h}_{(o1)A},$ $\tilde{h}_{(e0)},$
$\tilde{h}_{(e2)},$ $\tilde{h}_{(o2)}\}$ is not guaranteed.
For this reason, separate treatments for $l = 0,1$ modes are
required.

%*********************************************************************

To resolve this situation, in this paper, we introduce the mode
functions $k_{(\hat{\Delta})}$ and $k_{(\hat{\Delta}+2)m}$ instead of
$Y_{00}$ and $Y_{1m}$, respectively.
These mode functions satisfy the equations
\begin{eqnarray}
  \label{eq:kDetla-kDelta+2m-def}
  \hat{\Delta}k_{(\hat{\Delta})} = 0, \quad
  \left[\hat{\Delta}+2\right]k_{(\hat{\Delta}+2)m} = 0.
\end{eqnarray}
Although the tensor decomposition formula
(\ref{eq:hAB-fourier})--(\ref{eq:hpq-fourier}) with the harmonic
function $Y_{lm}$ does not have an inverse relation for the $l = 0,1$
modes, these equations may have an inverse relation even for the $l = 0,1$
modes if we choose the harmonic function $S$ such that
\begin{eqnarray}
  \label{eq:extended-harmonic-functions}
  S_{\delta} = \left\{
  \begin{array}{lcccl}
    Y_{lm} &\quad& \mbox{for} &\quad& l\geq 2; \\
    k_{(\hat{\Delta}+2)m} &\quad& \mbox{for} &\quad&  l=1; \\
    k_{(\hat{\Delta})} &\quad& \mbox{for} &\quad& l=0.
  \end{array}
                               \right.
\end{eqnarray}
Using Eq.~(\ref{eq:extended-harmonic-functions}) instead of $Y_{lm}$,
we expand the metric perturbation through the decomposition formulae
(\ref{eq:hAB-fourier})--(\ref{eq:hpq-fourier}).
To derive the inverse relation of this new decomposition formula, we
first use the fact that the operators $\hat{\Delta}^{-1}\hat{\Delta}$
and $[\hat{\Delta}+2]^{-1}[\hat{\Delta}+2]$ are projection operators,
which excludes the functions belonging to the kernels
${\cal K}_{\hat{\Delta}}:=\{f|\hat{\Delta} f=0\}$
and ${\cal K}_{\hat{\Delta}+2}:=\{f|[\hat{\Delta}+2]f=0\}$,
respectively.
For $l\geq 2$ mode, we use the orthogonality of spherical harmonics
$Y_{lm}$:
\begin{eqnarray}
  \int_{S^{2}} d\Omega Y_{lm}Y_{l'm'} = \delta_{ll'} \delta_{mm'}.
\end{eqnarray}
Furthermore, we can show that the set of the harmonic functions
\begin{eqnarray}
  \label{eq:set-of-harmonic-functions}
  \!\!\!\!\!\!\!\!\!\!\!\!\!\!\!\!\!\!\!\!
  \left\{S_{\delta}, \hat{D}_{p}S_{\delta},
  \epsilon_{pq}\hat{D}^{q}S_{\delta},
  \displaystyle \frac{1}{2} \gamma_{pq}S_{\delta},
  \left(\hat{D}_{p}\hat{D}_{q}-\frac{1}{2}\gamma_{pq}\hat{D}^{r}\hat{D}_{r}\right)S_{\delta},
  2\epsilon_{r(p}\hat{D}_{q)}\hat{D}^{r}S_{\delta}\right\}
\end{eqnarray}
are linear-independent if the conditions
\begin{eqnarray}
  &&
     \label{eq:condition-kernel-mode-of-Delta-def}
     k_{(\hat{\Delta})} \in {\cal K}_{\hat{\Delta}},
     \quad
     % \label{eq:condition-kernel-mode-of-Delta-cond}
     \hat{D}_{p}k_{(\hat{\Delta})} \neq 0,
     \quad
     % \label{eq:condition-kernel-mode-of-Delta-cond-2}
     \hat{D}_{p}\hat{D}_{q}k_{(\hat{\Delta})} \neq 0,
  \\
  &&
     \label{eq:condition-kernel-mode-of-Delta+2-def}
     k_{(\hat{\Delta}+2)m} \in {\cal K}_{(\hat{\Delta}+2)},
     \quad
     \hat{D}_{p}k_{(\hat{\Delta}+2)m} \neq 0,
     \quad
     % \label{eq:condition-kernel-mode-of-Delta+2-cond-1}
     k_{(\hat{\Delta}+2)m} = \Theta_{m}(\theta) e^{im\phi},
  \\
  &&
     \label{eq:condition-kernel-mode-of-Delta+2-cond-3}
     \left(\hat{D}_{p}\hat{D}_{q}k_{(\hat{\Delta}+2)m}\right)
     \left(\hat{D}^{p}\hat{D}^{q}k_{(\hat{\Delta}+2)m}\right)
     -
     2 \left(k_{(\hat{\Delta}+2)m}\right)^{2}
     \neq 0
\end{eqnarray}
are satisfied.
As the meaning of the conditions
(\ref{eq:condition-kernel-mode-of-Delta-def}), the set of
harmonic functions (\ref{eq:set-of-harmonic-functions}) is a
non-vanishing linear-independent set for $l = 0$ mode.
As the meaning of the conditions
(\ref{eq:condition-kernel-mode-of-Delta+2-def}) and
(\ref{eq:condition-kernel-mode-of-Delta+2-cond-3}), the set of
harmonic functions (\ref{eq:set-of-harmonic-functions}) are 
a non-vanishing linear-independent set for $l = 1$ mode.
The $\phi$-dependence of $k_{(\hat{\Delta}+2)m}$ in
Eq.~(\ref{eq:condition-kernel-mode-of-Delta+2-def}) is used to resolve
the $m=0,\pm 1$ mode-degeneracy through the formula
\begin{eqnarray}
  \label{eq:expimphi-orthogonality}
  \frac{1}{2\pi} \int_{0}^{2\pi} d\phi e^{i(m-m')\phi} = \delta_{mm'}.
\end{eqnarray}

%*********************************************************************

As the explicit functions of $k_{(\hat{\Delta})}$ and
$k_{(\hat{\Delta}+2)m}$, we employ the function
\begin{eqnarray}
  \label{eq:l=0-general-mode-func-specific}
  &&
     k_{(\hat{\Delta})}
     =
     1
     +
     {\cal \delta} \ln\left(\frac{1-z}{1+z}\right)^{1/2}
     \quad \delta\in\RF
     ,
  \\
  \label{eq:l=1-m=0-general-mode-func-specific}
  &&
     k_{(\hat{\Delta}+2)m=0}
     =
     z \left\{
     1
     +
     \delta
     \left(\frac{1}{2}\ln\frac{1+z}{1-z}-\frac{1}{z}\right)
     \right\}
     ,
  \\
  \label{eq:l=1-m=pm1-general-mode-func-specific}
  &&
     k_{(\hat{\Delta}+2)m=\pm 1}
     =
     (1-z^{2})^{1/2}
     \left\{
     1
     +
     \delta
     \left(\frac{1}{2}\ln\frac{1+z}{1-z}+\frac{z}{1-z^{2}}\right)
     \right\} e^{\pm i \phi}
     ,
\end{eqnarray}
where $z = \cos\theta$.
This choice satisfies the conditions
(\ref{eq:condition-kernel-mode-of-Delta-def})--(\ref{eq:condition-kernel-mode-of-Delta+2-cond-3}),
but is singular if $\delta\neq 0$.
When $\delta = 0$, we have $k_{(\hat{\Delta})}\propto Y_{00}$ and
$\hat{k}_{(\hat{\Delta}+2)m}\propto Y_{1m}$.

%*********************************************************************

Using the above harmonics functions $S_{\delta}$ in
Eq.~(\ref{eq:extended-harmonic-functions}), we propose the following strategy:
\begin{proposal}
  \label{proposal:harmonic-extension}
  We decompose the metric perturbations $h_{ab}$ on the background
  spacetime with the metric
  (\ref{eq:background-metric-2+2})--(\ref{eq:background-metric-2+2-gamma-comp-Schwarzschild}),
  through Eqs.~(\ref{eq:hAB-fourier})--(\ref{eq:hpq-fourier}) with the
  harmonic functions $S_{\delta}$ given by
  Eq.~(\ref{eq:extended-harmonic-functions}).
  Then, Eqs.~(\ref{eq:hAB-fourier})--(\ref{eq:hpq-fourier}) become
  invertible with the inclusion of $l=0,1$ modes.
  After deriving the field equations such as linearized Einstein
  equations using the harmonic function $S_{\delta}$, we choose
  $\delta=0$ when we solve these field equations as the regularity of
  the solutions.
\end{proposal}
Through this strategy, we can construct gauge-invariant variables and
evaluate the field equations through the mode-by-mode analyses
including $l = 0,1$ modes.

%*********************************************************************

%%%%%%%%%%%%%%%%%%%%%%%%%%%%%%%%%%%%%%%%%%%%%%%%%%%%%%
%%%%%%%%%%%%%%%%%%%%%%%%%%%%%%%%%%%%%%%%%%%%%%%%%%%%%%
%%%%%%%%%%%%%%%%%%%%%%%%%%%%%%%%%%%%%%%%%%%%%%%%%%%%%%
\section{Construction of gauge-invariant variables for linear perturbations}
\label{sec:Const._of_gauge_inv._var._linear_pert.}
%%%%%%%%%%%%%%%%%%%%%%%%%%%%%%%%%%%%%%%%%%%%%%%%%%%%%%
%%%%%%%%%%%%%%%%%%%%%%%%%%%%%%%%%%%%%%%%%%%%%%%%%%%%%%
%%%%%%%%%%%%%%%%%%%%%%%%%%%%%%%%%%%%%%%%%%%%%%%%%%%%%%

%*********************************************************************

Here, we consider the gauge-transformation rule
\begin{eqnarray}
  \label{eq:first-order-gauge-trans-of-metric-AIA2010-00100}
  {}_{\;{\cal Y}}\!h_{ab} - {}_{\;{\cal X}}\!h_{ab}
  =
  {\pounds}_{\xi}g_{ab}
  =
  2 \nabla_{(a}\xi_{b)}
\end{eqnarray}
for the linear-order perturbations on a spherically symmetric background
with the metric (\ref{eq:background-metric-2+2}).
We rewrite this gauge-transformation rule through the decomposition of
the generator
\begin{eqnarray}
  \label{eq:gauge-trans-generator-components}
  \xi_{a}
  &=&
      \xi_{A} (dx^{A})_{a} + \xi_{p} (dx^{p})_{a}
      ,
  \\
  \label{eq:gauge-trans-generator-xiA-xip-mode-sum}
  \xi_{A}
  &=&
      \sum_{l,m} \zeta_{A} S_{\delta}
      ,
      \quad
      \xi_{p}
      =
      r \sum_{l,m} \left(
      \zeta_{(e1)} \hat{D}_{p}S_{\delta}
      +
      \zeta_{(o1)} \epsilon_{pr}\hat{D}^{r}S_{\delta}
      \right)
      .
\end{eqnarray}
From the mode-decomposition
(\ref{eq:hAB-fourier})--(\ref{eq:hpq-fourier}), the mode-by-mode
components of the gauge-transformation rule
(\ref{eq:first-order-gauge-trans-of-metric-AIA2010-00100}) including
$l=0,1$ modes are summarized as follows:
\begin{eqnarray}
  \!\!\!\!\!\!\!\!\!\!\!\!
  &&
     {}_{{\cal Y}}\!\tilde{h}_{(o1)A}
     -
     {}_{{\cal X}}\!\tilde{h}_{(o1)A}
     =
     r
     \bar{D}_{A}
     \left(
     \frac{1}{r}
     \zeta_{(o1)}
     \right)
     ,
     \;\;\;
     {}_{{\cal Y}}\!\tilde{h}_{(o2)}
     -
     {}_{{\cal X}}\!\tilde{h}_{(o2)}
     =
     -
     \frac{1}{r}
     \zeta_{(o1)}
     ,
     \label{eq:gauge-trans-hAp-odd-lgeq2-sum}
  \\
  \!\!\!\!\!\!\!\!\!\!\!\!
  &&
     {}_{{\cal Y}}\tilde{h}_{AB}
     -
     {}_{{\cal X}}\tilde{h}_{AB}
     =
     2 \bar{D}_{(A}\zeta_{B)}
     ,
     \label{eq:gauge-trans-hAB-even-lgeq2-sum}
  \\
  \!\!\!\!\!\!\!\!\!\!\!\!
  &&
     {}_{{\cal Y}}\!\tilde{h}_{(e1)A}
     -
     {}_{{\cal X}}\!\tilde{h}_{(e1)A}
     =
     \frac{1}{r}
     \zeta_{A}
     +
     r
     \bar{D}_{A}
     \left(
     \frac{1}{r} \zeta_{(e1)}
     \right)
     ,
     \;\;\;
     {}_{{\cal Y}}\!\tilde{h}_{(e2)}
     -
     {}_{{\cal X}}\!\tilde{h}_{(e2)}
     =
     +
     \frac{2}{r}
     \zeta_{(e1)}
     ,
     \label{eq:gauge-trans-hAp-even-lgeq2-sum}
  \\
  \!\!\!\!\!\!\!\!\!\!\!\!
  &&
     {}_{{\cal Y}}\!\tilde{h}_{(e0)}
     -
     {}_{{\cal X}}\!\tilde{h}_{(e0)}
     =
     - \frac{2}{r} l(l+1) \zeta_{(e1)}
     +
     \frac{4}{r}
     \bar{D}^{A}r \zeta_{A}
     ,
     \label{eq:gauge-trans-hpq-trace-even-lgeq2-sum}
\end{eqnarray}
where $\bar{D}_{A}$ is the covariant derivative associated with the
metric $y_{AB}$ on ${\cal M}_{1}$.
The perturbations $\tilde{h}_{AB}$, $\tilde{h}_{(e1)A}$,
$\tilde{h}_{(e0)}$, and $\tilde{h}_{(e2)}$ and the generator
$\zeta_{A}$ and $\zeta_{(e1)}$ are called even modes, and the
perturbations $\tilde{h}_{(o1)A}$, $\tilde{h}_{(o2)}$, and
$\zeta_{(o1)}$ are called odd modes.
These even- and odd-mode perturbations are independent of each other,
and we may treat them separately.

%*********************************************************************

Inspecting gauge-transformation rules
(\ref{eq:gauge-trans-hAp-odd-lgeq2-sum}), for the odd mode,
gauge-invariant variable $\tilde{F}_{A}$ and gauge-variant variable
$\tilde{Y}_{(o2)}$ are defined by
\begin{eqnarray}
  \!\!\!\!\!\!\!\!\!\!\!\!
  \tilde{F}_{A}
  :=
  \tilde{h}_{(o1)A}
  + r \bar{D}_{A}\tilde{h}_{(o2)}
  ,
  \;\;\;
  \tilde{Y}_{(o2)}
  :=
  - r^{2} \tilde{h}_{(o2)}
  ,
  \;\;\;
  {}_{{\cal Y}}\tilde{Y}_{(o2)}
  -
  {}_{{\cal X}}\tilde{Y}_{(o2)}
  =
  r \zeta_{(o1)}
  .
  \label{eq:2+2-gauge-inv-odd-lgeq2-def-tildeFA}
\end{eqnarray}

%*********************************************************************

For even-mode perturbations, we first define the gauge-variant
variable $\tilde{Y}_{(e2)}$ by
\begin{eqnarray}
  \label{eq:gauge-variant-even-e1-def}
  \tilde{Y}_{(e2)}
  :=
  \frac{r^{2}}{2} \tilde{h}_{(e2)}
  ,
  \quad
  {}_{{\cal Y}}\!\tilde{Y}_{(e2)}
  -
  {}_{{\cal X}}\!\tilde{Y}_{(e2)}
  =
  r \zeta_{(e1)}
\end{eqnarray}
from the second equation in Eq.~(\ref{eq:gauge-trans-hAp-even-lgeq2-sum}).
Further, Eq.~(\ref{eq:gauge-trans-hAp-even-lgeq2-sum}) leads to
the following definition of $\tilde{Y}_{A}$:
\begin{eqnarray}
  \label{eq:2+2-gauge-trans-tildeYA-def}
  \tilde{Y}_{A}
  :=
  r \tilde{h}_{(e1)A}
  - \frac{r^{2}}{2} \bar{D}_{A}\tilde{h}_{(e2)}
  ,
  \quad
  {}_{{\cal Y}}\!\tilde{Y}_{A}
  -
  {}_{{\cal X}}\!\tilde{Y}_{A}
  =
  \zeta_{A}
  .
\end{eqnarray}
From the gauge-transformation rules
(\ref{eq:gauge-trans-hAB-even-lgeq2-sum}),
(\ref{eq:gauge-trans-hpq-trace-even-lgeq2-sum}),
(\ref{eq:gauge-variant-even-e1-def}), and
(\ref{eq:2+2-gauge-trans-tildeYA-def}), we define the two
gauge-invariant variables $\tilde{F}_{AB}$ and $\tilde{F}$ as
\begin{eqnarray}
  \label{eq:gauge-inv-lgeq2-tildeFAB-tildeF-def}
  \tilde{F}_{AB}
  :=
  \tilde{h}_{AB}
  - 2 \bar{D}_{(A}\tilde{Y}_{B)}
  ,
  \quad
  \tilde{F}
  :=
  \tilde{h}_{(e0)}
  - \frac{4}{r} \tilde{Y}_{A} \bar{D}^{A}r
  + \frac{2}{r^{2}} \tilde{Y}_{(e2)} l(l+1)
  .
\end{eqnarray}

%*********************************************************************

From the variables $\tilde{Y}_{(o2)}$, $\tilde{Y}_{(e2)}$, and
$\tilde{Y}_{A}$, which are defined by
Eqs.~(\ref{eq:2+2-gauge-inv-odd-lgeq2-def-tildeFA}),
(\ref{eq:gauge-variant-even-e1-def}), and
(\ref{eq:2+2-gauge-trans-tildeYA-def}), respectively,
we introduce the vector field $Y_{a}$ through
\begin{eqnarray}
  \label{eq:gauge-variant-vector-def}
  Y_{a}
  &:=&
       Y_{A} (dx^{A})_{a}
       +
       Y_{p} (dx^{p})_{a}
       ,
  \\
  \label{eq:gauge-variant-mode-sum-A-p}
  Y_{A}
  &:=&
       \sum_{l,m} \tilde{Y}_{A} S_{\delta}
       ,
       \quad
       Y_{p}
       :=
       \sum_{l,m} \left(
       \tilde{Y}_{(e2)} \hat{D}_{p}S_{\delta}
       +
       \tilde{Y}_{(o2)} \epsilon_{pr} \hat{D}^{r}S_{\delta}
       \right)
       .
\end{eqnarray}
Here, the gauge-transformation rule of the vector field $Y_{a}$ is
given by
\begin{eqnarray}
  \label{eq:gauge-trans-Ya}
  \!\!\!\!\!\!\!\!\!\!\!\!\!\!\!\!\!\!\!\!\!\!\!\!\!\!\!\!\!\!\!
  {}_{\cal Y}\!Y_{a} - {}_{\cal X}\!Y_{a}
  &=&
      \sum_{l,m} \zeta_{A} S_{\delta} (dx^{A})_{a}
      +
      r \sum_{l,m} \left(
      \zeta_{(e1)} \hat{D}_{p}S_{\delta}
      +
      \zeta_{(o1)} \epsilon_{pr} \hat{D}^{r}S_{\delta}
      \right)
      (dx^{p})_{a}
      =
      \xi_{a}
      .
\end{eqnarray}
We also introduce the gauge-invariant variables $F_{AB}$, $F_{Ap}$, and $F$
by
\begin{eqnarray}
  \label{eq:FAB-FAp-F-def}
  F_{AB}
  :=
  \sum_{l,m} \tilde{F}_{AB} S_{\delta}
  ,
  \quad
  F_{Ap}
  :=
  \sum_{l,m} \tilde{F}_{A} \epsilon_{pq} \hat{D}^{q}S_{\delta}
       ,
  \quad
  F
  :=
  \sum_{l,m} \tilde{F} S_{\delta}
  .
\end{eqnarray}
In terms of the variables
(\ref{eq:gauge-variant-mode-sum-A-p}) and
(\ref{eq:FAB-FAp-F-def}), the original components
(\ref{eq:hAB-fourier})--(\ref{eq:hpq-fourier}) of the metric
perturbations are given by
\begin{eqnarray}
  \label{eq:hAB-hAp-gauge-vari-inv-decomp}
  h_{AB}
  &=&
      F_{AB}
      +
      2
      \bar{D}_{(A}Y_{B)}
      ,
      \quad
      h_{Ap}
      =
      r
      F_{Ap}
      +
      \hat{D}_{p}Y_{A}
      +
      \bar{D}_{A}Y_{p}
      -
      \frac{2}{r} (\bar{D}_{A}r) Y_{p}
      ,
  \\
  \label{eq:hpq-gauge-vari-inv-decomp}
  h_{pq}
  &=&
      \frac{1}{2} \gamma_{pq}
      r^{2}
      F
      +
      2 r (\bar{D}^{A}r) \gamma_{pq} Y_{A}
      +
      2 \hat{D}_{(p}Y_{q)}
      .
\end{eqnarray}
The components of the gauge-invariant
metric perturbation ${\cal F}_{ab}$ are identified as
\begin{eqnarray}
  \label{eq:2+2-gauge-invariant-variables-calFAB-calFAp-calFpq}
  {\cal F}_{AB}
  :=
  F_{AB}
  ,
  \quad
  {\cal F}_{Ap}
  :=
  r F_{Ap}
  ,
  \quad
  {\cal F}_{pq}
  :=
  \frac{1}{2} \gamma_{pq} r^{2} F
  .
\end{eqnarray}
The expressions
(\ref{eq:hAB-hAp-gauge-vari-inv-decomp})--(\ref{eq:2+2-gauge-invariant-variables-calFAB-calFAp-calFpq})
are summarized as
\begin{eqnarray}
  \label{eq:2+2-decomposition-of-hab}
  h_{ab} = {\cal F}_{ab} + {\pounds}_{Y}g_{ab}.
\end{eqnarray}
Note that the above arguments include not only the $l\geq 2$ mode
but also $l=0,1$ modes of metric perturbations.
Equation (\ref{eq:2+2-decomposition-of-hab}) is a complete proof of the
Conjecture~\ref{conjecture:decomposition-conjecture} for the
perturbations on the spherically symmetric background spacetime.
Therefore, the general arguments in our gauge-invariant perturbation
theory are applicable to perturbations on the Schwarzschild background
spacetime without special treatment of $l=0,1$ modes.
We have therefore resolved the zero-mode problem in the perturbations on
the Schwarzschild background spacetime.

%*********************************************************************

%%%%%%%%%%%%%%%%%%%%%%%%%%%%%%%%%%%%%%%%%%%%%%%%%%%%%%
%%%%%%%%%%%%%%%%%%%%%%%%%%%%%%%%%%%%%%%%%%%%%%%%%%%%%%
%%%%%%%%%%%%%%%%%%%%%%%%%%%%%%%%%%%%%%%%%%%%%%%%%%%%%%
\section{$l=0,1$ solutions to the linearized Einstein equations}
\label{sec:l=01_sol_Einstein_equations}
%%%%%%%%%%%%%%%%%%%%%%%%%%%%%%%%%%%%%%%%%%%%%%%%%%%%%%
%%%%%%%%%%%%%%%%%%%%%%%%%%%%%%%%%%%%%%%%%%%%%%%%%%%%%%
%%%%%%%%%%%%%%%%%%%%%%%%%%%%%%%%%%%%%%%%%%%%%%%%%%%%%%

%*********************************************************************

As shown in Sec.~\ref{sec:review-of-perturbation-theroy}, the
linearized Einstein tensor ${}^{(1)}\!G_{a}^{\;\;b}$  for the linear
metric perturbation in the form (\ref{eq:2+2-decomposition-of-hab})
with the background Einstein equation $G_{a}^{\;\;b}=0$ is given by
${}^{(1)}\!G_{a}^{\;\;b}={}^{(1)}{\cal G}_{a}^{\;\;b}\left[{\cal F}\right]$
and
Eqs.~(\ref{eq:linear-Einstein-AIA2010-2})--(\ref{eq:(1)Sigma-def-linear}).
Herein, we consider the linearized non-vacuum Einstein equation
(\ref{eq:einstein-equation-gauge-inv}).
Because the background spacetime is the vacuum solution, the
first-order perturbations ${}^{(1)}\!T_{ac}$ and
${}^{(1)}\!T_{a}^{\;\;b}$ of the energy-momentum tensor are
automatically gauge-invariant as shown in
Eq.~(\ref{eq:Gab-Tab-decomp}).
We then have the gauge-invariant part of the first-order perturbation
of the energy-momentum tensor by ${}^{(1)}\!{\cal T}_{ac}$ $:=$
${}^{(1)}\!T_{ac}$ and ${}^{(1)}\!{\cal T}_{a}^{\;\;c}$ $:=$
${}^{(1)}\!T_{a}^{\;\;b}$ $=$ $g^{cb}{}^{(1)}\!{\cal T}_{ac}$.
Furthermore, because we only consider the perturbations on the vacuum
background solution based on the conventional general relativity, the
linear-metric perturbation $h_{ab}$ is not included in
${}^{(1)}\!{\cal T}_{ac}$ or ${}^{(1)}\!{\cal T}_{a}^{\;\;c}$.

%*********************************************************************

The total energy-momentum tensor satisfies the continuity equation
on the physical spacetime, which is pulled back to the background
spacetime.
Herein, we note that our background spacetime is the vacuum solution,
and the first-order perturbation of this continuity equation is given
by
Eq.~(\ref{eq:divergence-barTab-linear-vac-back-u})~\cite{K.Nakamura-2005}.
We decompose the linear-order perturbation of the energy-momentum
tensor ${}^{(1)}\!{\cal T}_{ac}$, and the components of this
tensor is given by
\begin{eqnarray}
  \!\!\!\!\!\!\!\!\!\!\!\!\!\!\!\!\!\!\!\!\!\!\!\!\!\!\!\!\!\!\!\!\!\!\!\!\!\!\!\!\!\!\!\!\!\!\!\!\!\!\!\!\!\!\!\!\!
  {}^{(1)}\!{\cal T}_{ac}
  &=&
      \sum_{l,m}
      \tilde{T}_{AC}
      S_{\delta}
      (dx^{A})_{a} (dx^{C})_{c}
      +
      2
      r
      \sum_{l,m} \left\{
      \tilde{T}_{(e1)A} \hat{D}_{p}S_{\delta}
      +
      \tilde{T}_{(o1)A} \epsilon_{pq} \hat{D}^{q}S_{\delta}
      \right\}
      (dx^{A})_{(a} (dx^{p})_{c)}
      \nonumber\\
  \!\!\!\!\!\!\!\!\!\!\!\!\!\!\!\!\!\!\!\!\!\!\!\!\!\!\!\!\!\!\!\!\!\!\!\!\!\!\!\!\!\!\!\!\!\!\!\!\!\!\!\!\!\!\!\!\!
  &&
     +
     r^{2}
     \sum_{l,m} \left\{
     \tilde{T}_{(e0)} \frac{1}{2} \gamma_{pq} S_{\delta}
     +
     \tilde{T}_{(e2)} \left(
     \hat{D}_{p}\hat{D}_{q}
     -
     \frac{1}{2} \gamma_{pq} \hat{D}_{r}\hat{D}^{r}
     \right) S_{\delta}
     \right.
     \nonumber\\
  \!\!\!\!\!\!\!\!\!\!\!\!\!\!\!\!\!\!\!\!\!\!\!\!\!\!\!\!\!\!\!\!\!\!\!\!\!\!\!\!\!\!\!\!\!\!\!\!\!\!\!\!\!\!\!\!\!
  && \quad\quad\quad\quad
     \left.
     +
     \tilde{T}_{(o2)} 2 \epsilon_{s(p}\hat{D}_{q)}\hat{D}^{s}S_{\delta}
     \right\}
     (dx^{p})_{(a} (dx^{q})_{c)}
     .
     \label{eq:1st-pert-calTab-dd-decomp-2}
\end{eqnarray}
In terms of these mode-decomposition, the components of the linearized
continuity equation (\ref{eq:divergence-barTab-linear-vac-back-u}) is
summarized as follows:
\begin{eqnarray}
  &&
     \bar{D}^{C}\tilde{T}_{C}^{\;\;B}
     + \frac{2}{r} (\bar{D}^{D}r)\tilde{T}_{D}^{\;\;\;B}
     -  \frac{1}{r} l(l+1) \tilde{T}_{(e1)}^{B}
     -  \frac{1}{r} (\bar{D}^{B}r) \tilde{T}_{(e0)}
     =
     0
     ,
     \label{eq:div-barTab-linear-vac-back-u-A-mode-dec-sum}
  \\
  &&
     \bar{D}^{C}\tilde{T}_{(e1)C}
     + \frac{3}{r} (\bar{D}^{C}r) \tilde{T}_{(e1)C}
     + \frac{1}{2r} \tilde{T}_{(e0)}
     -  \frac{1}{2r} (l-1)(l+2) \tilde{T}_{(e2)}
     =
     0
     ,
     \label{eq:div-barTab-linear-vac-back-u-p-mode-dec-even-sum}
  \\
  &&
     \bar{D}^{C}\tilde{T}_{(o1)C}
     + \frac{3}{r} (\bar{D}^{D}r) \tilde{T}_{(o1)D}
     + \frac{1}{r} (l-1)(l+2) \tilde{T}_{(o2)}
     =
     0
     .
     \label{eq:div-barTab-linear-vac-back-u-p-mode-dec-odd-sum}
\end{eqnarray}

%*********************************************************************

Now, we consider the solutions to the Einstein equation for $l=0,1$
mode imposing the regularity of the harmonics $S_{\delta}$ through the
choice $\delta=0$.
We should note that the harmonics
$\left(\hat{D}_{q}\hat{D}_{q}-\frac{1}{2}\gamma_{pq}\hat{\Delta}\right)S_{\delta}$
and $2\epsilon_{r(p}\hat{D}_{q)}\hat{D}^{r}S_{\delta}$ vanish for
$l=0,1$ modes when $\delta=0$.
This indicates that the components $\tilde{T}_{(e2)}$ and
$\tilde{T}_{(o2)}$ in Eq.~(\ref{eq:1st-pert-calTab-dd-decomp-2})
do not appear in the $l=0,1$ modes.
We may therefore safely choose $\tilde{T}_{(e2)}=0$ and
$\tilde{T}_{(o2)}=0$ for $l=1,0$ modes.
In addition, we also note that harmonics $\hat{D}_{p}S_{\delta}$ and
$\epsilon_{pq}\hat{D}^{q}S_{\delta}$ vanish for $l=0$ mode when
$\delta=0$.
This indicates that the components $\tilde{T}_{(e1)A}$ and
$\tilde{T}_{(o1)A}$ in Eq.~(\ref{eq:1st-pert-calTab-dd-decomp-2}) do
not appear in $l=0$ mode, and we may also choose
$\tilde{T}_{(e1)A}=0$ and $\tilde{T}_{(o1)A}=0$ for $l=0$ mode.
Through this choice and
Eq.~(\ref{eq:div-barTab-linear-vac-back-u-p-mode-dec-even-sum}), we
should regard $\tilde{T}_{(e0)}=0$ for $l=0$ mode.

%*********************************************************************

%%%%%%%%%%%%%%%%%%%%%%%%%%%%%%%%%%%%%%%%%%%%%%%%%%%%%%
%%%%%%%%%%%%%%%%%%%%%%%%%%%%%%%%%%%%%%%%%%%%%%%%%%%%%%
\subsection{$l=1$ odd mode perturbations}
\label{sec:Schwarzschild_Background-non-vacuum-Einstein-odd_l=1}
%%%%%%%%%%%%%%%%%%%%%%%%%%%%%%%%%%%%%%%%%%%%%%%%%%%%%%
%%%%%%%%%%%%%%%%%%%%%%%%%%%%%%%%%%%%%%%%%%%%%%%%%%%%%%

%*********************************************************************

If we impose the regularity on the harmonics $S_{\delta}$ by choosing
$\delta=0$, there is no $l=0$ mode in the odd-mode perturbation.
We therefore can concentrate only on $l=1$ mode.
Furthermore, we only consider $m=0$ mode because the
generalizations to $m=\pm 1$ modes are straightforward.
To evaluate the $l=1$ odd-mode solutions, we introduce the
components of $r \tilde{F}_{A}(dx^{A})_{a}$ by
\begin{eqnarray}
  \label{eq:component-odd-rtildeFD}
  r \tilde{F}_{A}(dx^{A})_{a} =: X_{(o)}(dt)_{a} + r^{2} \partial_{r}W_{(o)}(dr)_{a}.
\end{eqnarray}
Furthermore, it is convenient to introduce the functions $a_{1}(t,r)$ by
\begin{eqnarray}
  \frac{6M}{r^{4}} a_{1}(t,r)
  :=
  \partial_{r}
  \left(
  \frac{1}{r^{2}} X_{(o)}
  \right)
  -  \partial_{t}\partial_{r}W_{(o)}
  .
  \label{eq:A1-a1-variables-def}
\end{eqnarray}
Using the variables $a_{1}(t,r)$ and $rf \partial_{r}W_{(o)}$,
the Einstein equations for the $l=1$ odd-mode perturbations are
summarized as
\begin{eqnarray}
  &&
     \partial_{t}a_{1}(t,r)
     =
     -
     \frac{16 \pi}{3M} r^{3} f \tilde{T}_{(o1)r}
     ,
     \quad
     \partial_{r}a_{1}(t,r)
     =
     -
     \frac{16 \pi}{3M} \frac{r^{3}}{f} \tilde{T}_{(o1)t}
     ,
     \label{eq:a1-Einstein-eqs}
  \\
  &&
     \!\!\!\!\!\!\!\!\!\!\!\!\!\!\!\!\!\!\!
     + \partial_{t}^{2}(r f \partial_{r}W_{(o)})
     -  f \partial_{r}( f \partial_{r}(r f \partial_{r}W_{(o)})
     + \frac{1}{r^{2}} f \left[ 2 - 3 (1-f) \right] (r f \partial_{r}W_{(o)})
     \nonumber\\
  &=&
      16 \pi f \left(
      + f \tilde{T}_{(o1)r}
      + r f \bar{D}_{r}\tilde{T}_{(o2)}
      + (1-f) \tilde{T}_{(o2)}
      \right)
      .
      \label{eq:odd-master-equation-Regge-Wheeler-l=1-sum}
\end{eqnarray}
The integrability condition of
Eqs.~(\ref{eq:a1-Einstein-eqs}) is guaranteed by
Eq.~(\ref{eq:div-barTab-linear-vac-back-u-p-mode-dec-odd-sum}) with
$\tilde{T}_{(o2)}=0$.
Therefore, Eqs.~(\ref{eq:a1-Einstein-eqs}) are integrated as
\begin{eqnarray}
  a_{1}(t,r)
  &=&
      - \frac{16 \pi}{3M} r^{3} f \int dt \tilde{T}_{(o1)r} + a_{10}
      =
      - \frac{16 \pi}{3M} \int dr r^{3} \frac{1}{f} \tilde{T}_{(o1)t} + a_{10}
      \label{eq:a1tr-sol}
      ,
\end{eqnarray}
where $a_{10}$ is a constant.
By contrast,
Eq.~(\ref{eq:odd-master-equation-Regge-Wheeler-l=1-sum}) has the form
of the Regge-Wheeler equation~\cite{T.Regge-J.A.Wheeler-1957} with
$l=1$ for the variable $r f \partial_{r}W_{(o)}$, although the original
Regge-Wheeler equation is valid only in the case of $l\geq 2$.
From Eq.~(\ref{eq:a1tr-sol}) and the solution to
Eq.~(\ref{eq:odd-master-equation-Regge-Wheeler-l=1-sum}), the
component $X_{(o)}$ of $r\tilde{F}_{A}$ is obtained by the integration of
Eq.~(\ref{eq:A1-a1-variables-def}), and the explicit odd-mode solution
is given by
\begin{eqnarray}
  \label{eq:l=1-odd-mode-propagating-sol-ver2}
  &&
     \!\!\!\!\!\!\!\!\!\!\!\!\!\!\!\!\!\!\!\!\!\!\!\!\!\!\!\!\!\!\!\!\!\!\!\!
     2 {\cal F}_{Ap}(dx^{A})_{(a}(dx^{p})_{b)}
     =
     \left(
     6M
     r^{2} \int dr
     \frac{1}{r^{4}} a_{1}(t,r)
     \right) \sin^{2}\theta (dt)_{(a}(d\phi)_{b)}
     +
     {\pounds}_{V_{(o1)}}g_{ab}
     ,
  \\
  \label{eq:l=1-odd-mode-propagating-sol-ver2-Va-def}
  &&
     \!\!\!\!\!\!\!\!\!\!\!\!\!\!\!\!\!\!\!\!\!\!\!\!\!\!\!\!\!\!\!\!\!\!\!\!
     V_{(o1)a}
     =
     \left(\beta(t) + W_{(o)}(t,r)\right) r^{2} \sin^{2}\theta (d\phi)_{a}
     ,
\end{eqnarray}
where $\beta(t)$ is an arbitrary function of $t$.
In the vacuum case where $\tilde{T}_{(o1)r}=\tilde{T}_{(o1)t}=0$, the
function $a_{1}(t,r)$ becomes constant $a_{10}$.
In this case, (\ref{eq:l=1-odd-mode-propagating-sol-ver2}) is the
linearized Kerr solution with the Kerr parameter $a=a_{10}$ on the
Schwarzschild background spacetime, where $a$ is the total angular
momentum of the spacetime per mass~\cite{R.M.Wald-1984}.
Also note that the vector field $V_{(o1)a}$ and
${\pounds}_{V_{(o1)}}g_{ab}$ are gauge invariant.

%*********************************************************************

%%%%%%%%%%%%%%%%%%%%%%%%%%%%%%%%%%%%%%%%%%%%%%%%%%%%%%
%%%%%%%%%%%%%%%%%%%%%%%%%%%%%%%%%%%%%%%%%%%%%%%%%%%%%%
\subsection{$l=0,1$ even mode perturbations}
\label{sec:Schwarzschild_Background-non-vacuum-Einstein-even_l=01}
%%%%%%%%%%%%%%%%%%%%%%%%%%%%%%%%%%%%%%%%%%%%%%%%%%%%%%
%%%%%%%%%%%%%%%%%%%%%%%%%%%%%%%%%%%%%%%%%%%%%%%%%%%%%%

%*********************************************************************

Because the component $\tilde{T}_{(e2)}$ of the energy-momentum tensor
vanishes in both $l=0,1$ modes, one of the Einstein equations yields
\begin{eqnarray}
  \tilde{F}_{D}^{\;\;\;D} = 0.
  \label{eq:eventildeFtrace-matter-3}
\end{eqnarray}
We may then regard that the tensor $\tilde{F}_{AB}$ is traceless in
$l=0,1$ modes.
Furthermore, we introduce the components of $\tilde{F}_{AB}$ by
\begin{eqnarray}
  \label{eq:Xe-Ye-def-reduce}
  \!\!\!\!\!\!\!\!\!\!\!\!\!\!\!\!\!\!\!\!\!\!\!\!\!\!\!\!\!\!\!\!\!\!\!\!\!\!\!
  \tilde{F}_{AB}(dx^{A})_{a}(dx^{B})_{b}
  =:
  X_{(e)} \left\{ - f (dt)_{a}(dt)_{b} - f^{-1} (dr)_{a}(dr)_{b}\right\}.
  +
  2 Y_{(e)} (dt)_{(a}(dr)_{b)}
  .
\end{eqnarray}
In terms of the components $X_{(e)}$ and $Y_{(e)}$, the linearized
Einstein equations yield the initial value constraints
\begin{eqnarray}
  &&
     + \partial_{t}X_{(e)}
     + f \partial_{r}Y_{(e)}
     + \frac{1-f}{r} Y_{(e)}
     - \frac{1}{2} \partial_{t}\tilde{F}
     =
     16 \pi r \tilde{T}_{(e1)t}
     ,
     \label{eq:even-FAB-divergence-3-t-comp-sum-2}
  \\
  &&
     -  \partial_{t}Y_{(e)}
     -  f \partial_{r}X_{(e)}
     -  \frac{1-f}{r} X_{(e)}
     -
     \frac{1}{2} f \partial_{r}\tilde{F}
     =
     16 \pi r f \tilde{T}_{(e1)r}
     .
     \label{eq:even-FAB-divergence-3-r-comp-sum-2}
\end{eqnarray}
The other Einstein equations are three evolution equations for the
variables $X_{(e)}$, $Y_{(e)}$, and $\tilde{F}$.
To evaluate these evolution equations, it is convenient to introduce
the Moncrief variable $\Phi_{(e)}$ using
\begin{eqnarray}
  \label{eq:Moncrief-variable-def}
  \!\!\!\!\!\!\!\!
  \Phi_{(e)}
  :=
  \frac{r}{\Lambda} \left[
  f X_{(e)}
  - \frac{1}{4} \Lambda \tilde{F}
  + \frac{1}{2} r f \partial_{r}\tilde{F}
  \right]
  ,
  \;\;\;
  \Lambda := (l-1)(l+2)+3(1-f).
\end{eqnarray}
Furthermore, with these evolution equations and the above constraints
(\ref{eq:even-FAB-divergence-3-t-comp-sum-2}) and
(\ref{eq:even-FAB-divergence-3-r-comp-sum-2}), we have
\begin{eqnarray}
  \!\!\!\!\!\!\!\!\!\!\!\!\!\!\!\!\!
  l(l+1) Y_{(e)}
  &=&
      + \frac{2 \Lambda}{f}\partial_{t}\Phi_{(e)}
      + \frac{\Lambda+3f-1}{2f} r \partial_{t}\tilde{F}
     + 16 \pi r^{2} \tilde{T}_{tr}
     ,
     \label{eq:ll+1fYe-1st-pert-Ein-non-vac-tr-FAB-div-t-r-sum-2}
  \\
  \!\!\!\!\!\!\!\!\!\!\!\!\!\!\!\!\!
  l(l+1) \Lambda \tilde{F}
  &=&
      -  8 f \Lambda \partial_{r}\Phi_{(e)}
      + \frac{4}{r} \left[ 6 f (1-f) - l(l+1) \Lambda \right] \Phi_{(e)}.
      - 64 \pi r^{2} \tilde{T}_{tt}
      ,
      \label{eq:-Ein-non-vac-XF-constraint-Phie-reduce}
\end{eqnarray}
and the evolution equations
\begin{eqnarray}
  &&
  \!\!\!\!\!\!\!\!\!\!\!\!\!\!\!\!\!\!\!\!\!\!\!\!\!\!\!\!\!\!\!\!\!\!\!\!\!\!\!\!\!\!\!\!\!\!\!\!\!\!\!\!\!
  -  \frac{1}{f} \partial_{t}^{2}\Phi_{(e)}
  + \partial_{r}\left[ f \partial_{r}\Phi_{(e)} \right]
  -
  V_{even} \Phi_{(e)}
  =
  +  16 \pi \frac{r}{\Lambda} S_{(\Phi_{(e)})}
  ,
  \label{eq:Zerilli-Moncrief-eq-final-sum-reduce}
  \\
  &&
  \!\!\!\!\!\!\!\!\!\!\!\!\!\!\!\!\!\!\!\!\!\!\!\!\!\!\!\!\!\!\!\!\!\!\!\!\!\!\!\!\!\!\!\!\!\!\!\!\!\!\!\!\!
  V_{even}
  :=
  \frac{1}{r^{2}\Lambda^{2}}
  \left\{
  \Lambda^{2}
  \left[
  \Lambda - 2 (2-3f)
  \right]
  +
  6 (1-f)
  \left[
  (1-3f) \Lambda + 3 f (1-f)
  \right]
  \right\}
  ,
  \label{eq:Zerilli-Moncrief-master-potential-reduce}
  \\
  &&
  \!\!\!\!\!\!\!\!\!\!\!\!\!\!\!\!\!\!\!\!\!\!\!\!\!\!\!\!\!\!\!\!\!\!\!\!\!\!\!\!\!\!\!\!\!\!\!\!\!\!\!\!\!
  S_{(\Phi_{(e)})}
  :=
     \left( \frac{\Lambda}{4f} - \frac{1}{2} \right) \tilde{T}_{tt}
     -  \frac{1}{2} r \partial_{r}\tilde{T}_{tt}
     - \frac{3(1-f)}{\Lambda} \tilde{T}_{tt}
     + \left( \frac{2-f}{2} - \frac{\Lambda}{4} \right) f \tilde{T}_{rr}
     + \frac{1}{2} f^{2} r \partial_{r}\tilde{T}_{rr}
     \nonumber\\
  &&
  \!\!\!\!\!\!\!\!\!\!\!\!\!\!\!\!\!\!\!\!\!
     -  \frac{f}{2} \tilde{T}_{(e0)}
     -  l(l+1) f \tilde{T}_{(e1)r}
     .
     \label{eq:SPhie-def-explicit-reduce}
\end{eqnarray}
Equation (\ref{eq:Zerilli-Moncrief-eq-final-sum-reduce}) is the
Zerilli equation.
Furthermore, the evolution equation for $\tilde{F}$ given by
\begin{eqnarray}
  \!\!\!\!\!\!\!\!\!\!\!\!\!\!\!\!\!\!\!\!\!\!\!\!\!\!\!\!\!\!\!\!\!\!\!\!\!\!\!\!\!\!\!\!\!\!\!\!\!\!\!\!\!
  -  \frac{1}{f} \partial_{t}^{2}\tilde{F}
  + \partial_{r}( f \partial_{r}\tilde{F} )
  + \frac{1}{r^{2}} 3(1-f) \tilde{F}
  + \frac{4}{r^{3}} \Lambda \Phi_{(e)}
  =
  16 \pi \left[
  -  \frac{1}{f} \tilde{T}_{tt}
  + f \tilde{T}_{rr}
  + 4 f \tilde{T}_{(e1)r}
  \right]
  \label{eq:even-mode-tildeF-eq-Phie-reduce}
\end{eqnarray}
is also useful.
We can check the consistency of
Eqs.~(\ref{eq:even-FAB-divergence-3-t-comp-sum-2})--(\ref{eq:even-mode-tildeF-eq-Phie-reduce})
based on the continuity equations
(\ref{eq:div-barTab-linear-vac-back-u-A-mode-dec-sum}),
(\ref{eq:div-barTab-linear-vac-back-u-p-mode-dec-even-sum}), and
$\tilde{T}_{(e2)}=0$ of the energy-momentum tensor.

%*********************************************************************

%%%%%%%%%%%%%%%%%%%%%%%%%%%%%%%%%%%%%%%%%%%%%%%%%%%%%%
\subsubsection{$l=0$ mode solutions}
\label{sec:non-vacuum-Einstein-even_l=0_sol}
%%%%%%%%%%%%%%%%%%%%%%%%%%%%%%%%%%%%%%%%%%%%%%%%%%%%%%

%*********************************************************************

In the $l=0$ case,
Eqs.~(\ref{eq:ll+1fYe-1st-pert-Ein-non-vac-tr-FAB-div-t-r-sum-2}) and
(\ref{eq:-Ein-non-vac-XF-constraint-Phie-reduce}) do not yield the
variables $Y_{(e)}$ or $\tilde{F}$ as solutions.
Instead, introducing the variable
\begin{eqnarray}
  \label{eq:m1tr-def}
  m_{1}(t,r) := - \frac{1}{2} (1 - 3f) \Phi_{(e)},
\end{eqnarray}
which corresponds to the mass perturbations, these equations are given by
\begin{eqnarray}
  \partial_{t}m_{1}(t,r)
  =
  4 \pi r^{2} f \tilde{T}_{tr}
  , \quad
  \partial_{r}m_{1}(t,r)
  =
  4 \pi r^{2} \frac{1}{f} \tilde{T}_{tt}
  .
  \label{eq:Ein-non-vac-XF-constraint-Phie-l=0}
\end{eqnarray}
The integrability of
Eqs.~(\ref{eq:Ein-non-vac-XF-constraint-Phie-l=0}) is guaranteed
through the $t$-component of the continuity equation
(\ref{eq:div-barTab-linear-vac-back-u-A-mode-dec-sum}).
Furthermore, the $l=0$ mode version of
Eq.~(\ref{eq:Zerilli-Moncrief-eq-final-sum-reduce}) with the potential
(\ref{eq:Zerilli-Moncrief-master-potential-reduce}) and the source
term (\ref{eq:SPhie-def-explicit-reduce}) is trivial.
Actually, Eqs.~(\ref{eq:Ein-non-vac-XF-constraint-Phie-l=0}) are
integrated as
\begin{eqnarray}
  \!\!\!\!\!\!\!\!
  m_{1}(t,r)
  =
  4 \pi \int dr \left[\frac{r^{2}}{f} \tilde{T}_{tt}\right]
  + M_{1}
  =
  4 \pi \int dt \left[ r^{2} f \tilde{T}_{rt} \right]
  + M_{1}
  ,
  \quad
  M_{1}\in\RF
  .
  \label{eq:Ein-non-vac-m1-sol.}
\end{eqnarray}
The solution (\ref{eq:Ein-non-vac-m1-sol.}) to
Eqs.~(\ref{eq:Ein-non-vac-XF-constraint-Phie-l=0}) gives the variable
$\Phi_{(e)}$ as a solution to the Einstein equation through
Eq.~(\ref{eq:m1tr-def}).
The variable $\tilde{F}$ is obtained as a solution to
Eq.~(\ref{eq:even-mode-tildeF-eq-Phie-reduce}) with the solution
$\Phi_{(e)}$.
From $(\tilde{F},\Phi_{(e)})$, the variable
$X_{(e)}$ is given as a solution to the Einstein equation through
Eq.~(\ref{eq:Moncrief-variable-def}).
Through the solution $(\tilde{F},X_{(e)})$, the variable $Y_{(e)}$ is
obtained from the constraints
(\ref{eq:even-FAB-divergence-3-t-comp-sum-2}) and
(\ref{eq:even-FAB-divergence-3-r-comp-sum-2}).

%*********************************************************************

Note that Eq.~(\ref{eq:even-mode-tildeF-eq-Phie-reduce}) has the same
form as the inhomogeneous version of the Regge-Wheeler equation with
$l=0$, although the original Regge-Wheeler equation is valid only in
the case $l\geq 2$.
We denote the solution to
Eq.~(\ref{eq:even-mode-tildeF-eq-Phie-reduce}) as
\begin{eqnarray}
  \label{eq:tildeF-sol.}
  \tilde{F} =: \partial_{t}\Upsilon.
\end{eqnarray}
Using this solution (\ref{eq:tildeF-sol.}), we obtain
\begin{eqnarray}
  f X_{(e)}
  =
  -  \frac{2m_{1}(t,r)}{r}
  + \frac{1}{4} (1-3f) \partial_{t}\Upsilon
  -  \frac{1}{2} r f \partial_{r}\partial_{t}\Upsilon
  .
  \label{eq:l=0-version-non-vac-Xe-sol}
\end{eqnarray}
The integrability condition of
Eqs.~(\ref{eq:even-FAB-divergence-3-t-comp-sum-2}) and
(\ref{eq:even-FAB-divergence-3-r-comp-sum-2}) for the $l=0$ mode is
guaranteed through
Eqs.~(\ref{eq:Zerilli-Moncrief-eq-final-sum-reduce}) and
(\ref{eq:even-mode-tildeF-eq-Phie-reduce}), and
Eqs.~(\ref{eq:even-FAB-divergence-3-t-comp-sum-2}) and
(\ref{eq:even-FAB-divergence-3-r-comp-sum-2}) are integrated as
follows:
\begin{eqnarray}
  Y_{(e)}
  &=&
      -  \frac{2}{r^{2}} \int dt m_{1}(t,r)
      + \frac{3}{4r} (1-f) \Upsilon
      -  \frac{1}{4} (1-3f) \partial_{r}\Upsilon
      + \frac{1}{2} r \partial_{r}( f \partial_{r}\Upsilon )
      \nonumber\\
  &&
     + 8 \pi \frac{r}{f} \int dt \tilde{T}_{tt}
     + \frac{1}{2} r  \zeta(r)
     -  \frac{1}{4f} \int dr (1-3f) \zeta(r)
     + \frac{\xi}{f}
     ,
     \label{eq:even-FAB-divergence-3-r-t-sol}
\end{eqnarray}
where $\zeta(r)$ is an arbitrary function of $r$, and $\xi$ is an
arbitrary constant.
Substituting Eqs.~(\ref{eq:l=0-version-non-vac-Xe-sol}) and
(\ref{eq:even-FAB-divergence-3-r-t-sol}) into
(\ref{eq:Xe-Ye-def-reduce}) using (\ref{eq:FAB-FAp-F-def}), and
(\ref{eq:2+2-gauge-invariant-variables-calFAB-calFAp-calFpq}), we
obtain
\begin{eqnarray}
  {\cal F}_{ab}
  &=&
      \frac{2}{r}
      \left(
      M_{1}
      + 4 \pi \int dr \left[\frac{r^{2}}{f} \tilde{T}_{tt}\right]
      \right)
      \left(
      (dt)_{a}(dt)_{b}
      +
      \frac{1}{f^{2}}
      (dr)_{a}(dr)_{b}
      \right)
      \nonumber\\
  &&
     +
     2 \left[
     4 \pi r \int dt \left(
     \frac{1}{f} \tilde{T}_{tt}
     + f \tilde{T}_{rr}
     \right)
     \right]
     (dt)_{(a}(dr)_{b)}
      \nonumber\\
  &&
     +
     {\pounds}_{V_{(e0)}}g_{ab}
     ,
     \label{eq:l=0-final-sols}
\end{eqnarray}
where we defined the vector field $V_{(e0)a}$ and an arbitrary function
$\gamma(r)$ by
\begin{eqnarray}
  \label{eq:Va-second-choice-non-vac-sum}
  V_{(e0)a}
  &:=&
      \left(
      \frac{1}{4} f \Upsilon + \frac{1}{4} r f \partial_{r}\Upsilon
      + \gamma(r)
      \right)
      (dt)_{a}
      +
      \frac{1}{4f} r \partial_{t}\Upsilon
      (dr)_{a}
      ,
  \\
  \label{eq:Va-second-choice-gamma-non-vac-sum}
  \gamma(r)
  &:=&
      f \int dr \left[
      + \frac{1}{4f} r \zeta(r)
      -  \frac{1}{4f^{2}} \int dr (1-3f) \zeta(r)
      + \frac{\xi}{f^{2}}
      \right]
      .
\end{eqnarray}
Herein, we note that the vector field $V_{a}$ defined by
Eq.~(\ref{eq:Va-second-choice-non-vac-sum}) is gauge invariant.

%*********************************************************************

%%%%%%%%%%%%%%%%%%%%%%%%%%%%%%%%%%%%%%%%%%%
\subsubsection{$l=1$ mode solutions}
\label{sec:l=1_Schwarzschild_Background-non-vac}
%%%%%%%%%%%%%%%%%%%%%%%%%%%%%%%%%%%%%%%%%%%

%*********************************************************************

In the $l=1$ even-mode case, we obtain the components $\tilde{F}$,
$Y_{(e)}$, and $X_{(e)}$ through
Eqs.~(\ref{eq:-Ein-non-vac-XF-constraint-Phie-reduce}),
(\ref{eq:ll+1fYe-1st-pert-Ein-non-vac-tr-FAB-div-t-r-sum-2}), and
(\ref{eq:Moncrief-variable-def}) if we obtain the
variable $\Phi_{(e)}$ as a solution to the linearized Einstein equations.
The variable $\Phi_{(e)}$ is determined by the master equation
(\ref{eq:Zerilli-Moncrief-eq-final-sum-reduce}) with the appropriate
boundary conditions.
From Eqs.~(\ref{eq:eventildeFtrace-matter-3}),
(\ref{eq:Xe-Ye-def-reduce}), (\ref{eq:FAB-FAp-F-def}), and
(\ref{eq:2+2-gauge-invariant-variables-calFAB-calFAp-calFpq}), we
obtain the gauge-invariant metric perturbation ${\cal F}_{ab}$.
Herein, we only consider the $m=0$ mode in the $l=1$ mode perturbations.
We then obtain
\begin{eqnarray}
  {\cal F}_{ab}
  &=&
      - \frac{16\pi r^{2} f^{2}}{3(1-f)} \left[
      \frac{1+f}{2} \tilde{T}_{rr}
      + r f \partial_{r}\tilde{T}_{rr}
      -  \tilde{T}_{(e0)}
      -  4 \tilde{T}_{(e1)r}
      \right] \cos\theta (dt)_{a}(dt)_{b}
      \nonumber\\
  &&
      +
      16 \pi r^{2} \left\{
      \tilde{T}_{tr}
      - \frac{2r}{3f(1-f)} \partial_{t}\tilde{T}_{tt}
      \right\} \cos\theta (dt)_{(a}(dr)_{b)}
      \nonumber\\
  &&
     + \frac{8 \pi r^{2} (1-3f)}{f^{2}(1-f)} \left[
     \tilde{T}_{tt}
     -  \frac{2rf}{3(1-3f)} \partial_{r}\tilde{T}_{tt}
     \right] \cos\theta (dr)_{a}(dr)_{b}
     \nonumber\\
  &&
      -  \frac{16 \pi r^{4}}{3(1-f)} \tilde{T}_{tt} \cos\theta \gamma_{ab}
     +
     {\pounds}_{V_{(e1)}}g_{ab}
      ,
      \label{eq:calFab-l=1-m=0-sol.}
\end{eqnarray}
where
\begin{eqnarray}
  V_{(e1)a}
  &:=&
       -  r \partial_{t}\Phi_{(e)} \cos\theta (dt)_{a}
       + \left( \Phi_{(e)} - r \partial_{r}\Phi_{(e)} \right) \cos\theta (dr)_{a}
       \nonumber\\
  &&
     -  r \Phi_{(e)} \sin\theta (d\theta)_{a}
     .
     \label{eq:generator-covariant-l=1-m=0-sum-2}
\end{eqnarray}
Here, the vector field $V_{(e1)a}$ is also gauge invariant.
The components of the energy-momentum tensor in
Eq.~(\ref{eq:calFab-l=1-m=0-sol.}) satisfy the linear perturbations
(\ref{eq:div-barTab-linear-vac-back-u-A-mode-dec-sum}) and
(\ref{eq:div-barTab-linear-vac-back-u-p-mode-dec-even-sum}).
We also note that there may exist an additional gauge-invariant term
that has the form of the Lie derivative of the background metric in
addition to the term ${\pounds}_{V_{(e1)}}g_{ab}$ in
Eq.~(\ref{eq:calFab-l=1-m=0-sol.}).
This depends on the equation of state of the perturbation of the
energy-momentum tensor.

%*********************************************************************

%%%%%%%%%%%%%%%%%%%%%%%%%%%%%%%%%%%%%%%%%%%%%%%%%%%%%%
%%%%%%%%%%%%%%%%%%%%%%%%%%%%%%%%%%%%%%%%%%%%%%%%%%%%%%
%%%%%%%%%%%%%%%%%%%%%%%%%%%%%%%%%%%%%%%%%%%%%%%%%%%%%%
\section{Summary and Discussions}
\label{sec:summary_and_discussion}
%%%%%%%%%%%%%%%%%%%%%%%%%%%%%%%%%%%%%%%%%%%%%%%%%%%%%%
%%%%%%%%%%%%%%%%%%%%%%%%%%%%%%%%%%%%%%%%%%%%%%%%%%%%%%
%%%%%%%%%%%%%%%%%%%%%%%%%%%%%%%%%%%%%%%%%%%%%%%%%%%%%%

%****************************************************************

To summarize, we proposed a gauge-invariant treatment of the
$l=0,1$-mode perturbations on the Schwarzschild background spacetime.
Instead of the spherical harmonics $Y_{lm}$ with $l=0,1$, we used the
mode functions $k_{(\hat{\Delta})}$ and $k_{(\hat{\Delta}+2)m}$ with the
parameter $\delta$.
These functions are the kernel mode of the derivative operators
$\hat{\Delta}$ and $\hat{\Delta}+2$, respectively.
We chose the parameter $\delta$ such that the choice $\delta=0$
realizes the usual spherical harmonics $Y_{lm}$, and examined the
linear independence of the scalar harmonic functions $S_{\delta}$
defined by (\ref{eq:extended-harmonic-functions}),
the vector harmonics $\hat{D}_{p}S_{\delta}$ and
$\epsilon_{pr}\hat{D}^{r}S_{\delta}$, and the tensor harmonics
$\frac{1}{2}\gamma_{pq}S_{\delta}$,
$\left(\hat{D}_{p}\hat{D}_{q}-\frac{1}{2}\gamma_{pq}\hat{D}^{r}\hat{D}_{r}\right)S_{\delta}$,
and $2\epsilon_{r(p}\hat{D}_{q)}\hat{D}^{r}S_{\delta}$.
We thus proposed Proposal~\ref{proposal:harmonic-extension} as a
strategy of a gauge-invariant treatment of the $l=0,1$ perturbations
on the Schwarzschild background spacetime.
Following this proposal, we derived the $l=0,1$ mode solutions to the
Einstein equations with the general linear perturbations of the
energy-momentum tensor in the gauge-invariant manner.
Herein, it is assumed that these general linear perturbations of the
energy-momentum tensor satisfy the linear perturbations of the
divergence of the energy-momentum tensor.

%****************************************************************

The derived solution in the $l=1$ odd mode actually realizes the
linearized Kerr solution in the vacuum case.
Apart from the term described as the Lie derivative of the
background metric, the unique solution in the odd-mode vacuum
perturbation case is the Kerr parameter perturbation.
Furthermore, we also derived the $l=0,1$ even-mode solutions to the
Einstein equations.
In the vacuum case, in which all components of ${}^{(1)}\!{\cal T}_{ab}$
vanish, a $l=0$ even-mode solution realizes the only the additional
mass parameter perturbation of the Schwarzschild spacetime, apart from
the terms described by the Lie derivative of the background metric.
This is the realization of the linearized gauge-invariant version of
Birkhoff's theorem.
Owing to this realization, we can state that our proposal is physically
reasonable.
Herein, we note that the terms described by the Lie derivative of the
background spacetime in Eq.~(\ref{eq:l=0-final-sols}) is necessary if
we include the Schwarzschild mass perturbation $M_{1}$ as the solution
to the linearized Einstein equations.
Actually, if we choose $\Upsilon=0$, the mass parameter $M_{1}$ must
vanish from Eq.~(\ref{eq:even-mode-tildeF-eq-Phie-reduce}).

%****************************************************************

Also note that all $l=1$ odd-mode and $l=0,1$ even-mode
solutions include the term which is described by the Lie derivative of
the background metric.
Because the definitions of gauge-invariant and gauge-variant variables
are not unique, as explained through Eq.~(\ref{eq:gauge-inv-nonunique})
in Sec.~\ref{sec:review-of-perturbation-theroy}, such terms may appear
in our perturbation theory.
Furthermore, although these terms can be eliminated through the
gauge-fixing method at any time, they are gauge-invariant, i.e., they
are invariant under the change of point-identifications between the
physical spacetime ${\cal M}_{\rm ph}$ and the background spacetime
${\cal M}$.
The gauge-invariance of these terms implies that the
point-identification between the physical spacetime
${\cal M}_{\rm ph}$ and the background spacetime ${\cal M}$ are
already fixed.
Therefore, these terms may have a physical meaning.
As an example, the function $\beta(t)$ in
Eq.~(\ref{eq:l=1-odd-mode-propagating-sol-ver2-Va-def}) can be
produced by the infinitesimal coordinate transformation
$\phi\rightarrow\phi+\omega(t)t$ in the background metric
(\ref{eq:background-metric-2+2}) with
Eqs.~(\ref{eq:background-metric-2+2-y-comp-Schwarzschild}) and (\ref{eq:background-metric-2+2-gamma-comp-Schwarzschild}).
Because this function $\beta(t)$ is gauge-invariant, the coordinate
transformation $\phi\rightarrow\phi+\omega(t)t$ should not be regard
as the coordinate transformation (\ref{eq:induced-coordinate-trans})
induced by the gauge-transformation $\Phi_{\lambda}$ described in
Sec.~\ref{sec:review-of-perturbation-theroy} but should be
regarded as the coordinate transformation within the background
spacetime, i.e., the first-kind gauge transformation on the background
spacetime, which is commented in
Sec.~\ref{sec:review-of-perturbation-theroy}.
Because the point-identification is already fixed owing to the
gauge-invariance of this term, the coordinate transformation on
the background spacetime ${\cal M}$ is also regarded as the coordinate
transformation on the physical spacetime ${\cal M}_{{\rm ph}}$.
Because the coordinate transformation $\phi\rightarrow\phi+\omega(t)t$
describes the rotation of the Universe, it is regarded as the
coordinate transformation into a non-inertia frame and the function
$\beta(t)$ represents an inertia force which appears as a property of
the physical spacetime ${\cal M}_{{\rm ph}}$.
This will be a physical meaning of the function $\beta(t)$ in
Eq.~(\ref{eq:l=1-odd-mode-propagating-sol-ver2-Va-def}).
Of course, it is highly non-trivial that all terms in the $l=1$ odd-mode
and $l=0,1$ even-mode solutions that have the form
of the Lie derivative of the background metric can be interpreted in
a similar manner as this function $\beta(t)$, which remains
an open question.

%****************************************************************

Finally, we should emphasize that we confirmed
Conjecture~\ref{conjecture:decomposition-conjecture} for the
linear-metric perturbations in the Schwarzschild background case,
including the $l=0,1$ modes.
Because these are zero modes in Refs.~\cite{K.Nakamura-2011}, we
resolved the zero-mode problem for the perturbations on the Schwarzschild
background spacetime.
Conjecture~\ref{conjecture:decomposition-conjecture} is important
and is the only non-trivial premise of our general framework of the
gauge-invariant higher-order perturbation theory.
For this reason, in principle, the extension to any-order
perturbations through our gauge-invariant
formulation~\cite{K.Nakamura-2014} is possible, at least in the
Schwarzschild background case.
Although a short discussion was already given in our
companion paper~\cite{K.Nakamura-2021b},
we will discuss this extension to the higher-order perturbation
elsewhere.

%*******************************************************************

%%%%%%%%%%%%%%%%%%%%%%%%%%%%%%%%%%%%%%%%%%%%%%%%%%%%%%
%%%%%%%%%%%%%%%%%%%%%%%%%%%%%%%%%%%%%%%%%%%%%%%%%%%%%%
%%%%%%%%%%%%%%%%%%%%%%%%%%%%%%%%%%%%%%%%%%%%%%%%%%%%%%
%\appendix
%%%%%%%%%%%%%%%%%%%%%%%%%%%%%%%%%%%%%%%%%%%%%%%%%%%%%%
%%%%%%%%%%%%%%%%%%%%%%%%%%%%%%%%%%%%%%%%%%%%%%%%%%%%%%
%%%%%%%%%%%%%%%%%%%%%%%%%%%%%%%%%%%%%%%%%%%%%%%%%%%%%%

%*******************************************************************

%%%%%%%%%%%%%%%%%%%%%%%%%%%%%%%%%%%%%%%%%%%%%%%%%%%%%%
%%%%%%%%%%%%%%%%%%%%%%%%%%%%%%%%%%%%%%%%%%%%%%%%%%%%%%
%%%%%%%%%%%%%%%%%%%%%%%%%%%%%%%%%%%%%%%%%%%%%%%%%%%%%%
\section*{Acknowledgements}
%%%%%%%%%%%%%%%%%%%%%%%%%%%%%%%%%%%%%%%%%%%%%%%%%%%%%%
%%%%%%%%%%%%%%%%%%%%%%%%%%%%%%%%%%%%%%%%%%%%%%%%%%%%%%
%%%%%%%%%%%%%%%%%%%%%%%%%%%%%%%%%%%%%%%%%%%%%%%%%%%%%%

%****************************************************************

The author would like to thank Prof. Shuhei Mano for the valuable comments and discussions.
The author also deeply acknowledges Prof. Hiroyuki Nakano for various
discussions and suggestions during the past 20 years.

%****************************************************************

%%%%%%%%%%%%%%%%%%%%%%%%%%%%%%%%%%%%%%%%%%%%%%%%%%%%%%
%%%%%%%%%%%%%%%%%%%%%%%%%%%%%%%%%%%%%%%%%%%%%%%%%%%%%%
%%%%%%%%%%%%%%%%%%%%%%%%%%%%%%%%%%%%%%%%%%%%%%%%%%%%%%
\section*{References}
%%%%%%%%%%%%%%%%%%%%%%%%%%%%%%%%%%%%%%%%%%%%%%%%%%%%%%
%%%%%%%%%%%%%%%%%%%%%%%%%%%%%%%%%%%%%%%%%%%%%%%%%%%%%%
%%%%%%%%%%%%%%%%%%%%%%%%%%%%%%%%%%%%%%%%%%%%%%%%%%%%%%

\end{document}